\documentclass[aps,prd
,preprint,tightenlines,superscriptaddress,nofootinbib,showpacs]{revtex4}
\usepackage{amssymb,latexsym}
\usepackage{amsmath,amsbsy,bbm}
\usepackage{epsfig,bm}
\usepackage{graphicx,comment}
\usepackage{color}
\DeclareMathOperator{\tr}{tr}

\begin{document}
\def\a{{\alpha}}
\def\b{{\beta}}
\def\d{{\delta}}
\def\D{{\Delta}}
\def\e{{\varepsilon}}
\def\g{{\gamma}}
\def\G{{\Gamma}}
\def\k{{\kappa}}
\def\l{{\lambda}}
\def\L{{\Lambda}}
\def\m{{\mu}}
\def\n{{\nu}}
\def\o{{\omega}}
\def\O{{\Omega}}
\def\S{{\Sigma}}
\def\s{{\sigma}}
\def\th{{\theta}}
\newcommand{\mnod}{\stackrel{\circ}{M}}

\def\ol#1{{\overline{#1}}}

\def\Dslash{\ol D\hskip-0.65em /}
\def\Dslashe{D\hskip-0.65em /}
\def\Pslash{\ol P\hskip-0.65em /}
\def\lslash{l\hskip-0.35em /}
\def\Pslashe{P\hskip-0.65em /}

\def\Dtslash{\tilde{D} \hskip-0.65em /}

\def\CPT{{$\chi$PT}}
\def\QCPT{{Q$\chi$PT}}
\def\PQCPT{{PQ$\chi$PT}}
\def\tr{\text{tr}}
\def\str{\text{str}}
\def\diag{\text{diag}}
\def\order{{\mathcal O}}

\def\cC{{\mathcal C}}
\def\cB{{\mathcal B}}
\def\cT{{\mathcal T}}
\def\cQ{{\mathcal Q}}
\def\cL{{\mathcal L}}
\def\cO{{\mathcal O}}
\def\cA{{\mathcal A}}
\def\cQ{{\mathcal Q}}
\def\cR{{\mathcal R}}
\def\cH{{\mathcal H}}
\def\cW{{\mathcal W}}
\def\cM{{\mathcal M}}
\def\cD{{\mathcal D}}
\def\cN{{\mathcal N}}
\def\cP{{\mathcal P}}
\def\cK{{\mathcal K}}
\def\Qt{{\tilde{Q}}}
\def\Dt{{\tilde{D}}}
\def\St{{\tilde{\Sigma}}}
\def\cBt{{\tilde{\mathcal{B}}}}
\def\cDt{{\tilde{\mathcal{D}}}}
\def\cTt{{\tilde{\mathcal{T}}}}
\def\cMt{{\tilde{\mathcal{M}}}}
\def\At{{\tilde{A}}}
\def\cNt{{\tilde{\mathcal{N}}}}
\def\cOt{{\tilde{\mathcal{O}}}}
\def\cPt{{\tilde{\mathcal{P}}}}
\def\cI{{\mathcal{I}}}
\def\cJ{{\mathcal{J}}}
\def\cb{{\cal B}}
\def\cbb{{\overline{\cal B}}}

\def\eqref#1{{(\ref{#1})}}

\preprint{UMD-40762-405}
 
\title{Chiral Corrections to Hyperon Axial Form Factors}

\author{Fu-Jiun~Jiang}
\email[]{fjjiang@itp.unibe.ch}
\affiliation{Institute for Theoretical Physics, Bern University, Sidlerstrasse 5, CH-3012 Bern, Switzerland}
\author{B.~C.~Tiburzi}
\email[]{bctiburz@umd.edu}
\affiliation{Department of Physics, Duke University, Box 90305, Durham, NC 27708-0305, USA}
\affiliation{Maryland Center for Fundamental Physics, Department of Physics, University of Maryland, College Park, MD 20742-4111, USA}

\date{\today}

\pacs{12.38.Gc, 12.39.Fe}

\begin{abstract}
We study the complete set of flavor changing hyperon axial current matrix 
elements
at small momentum transfer. Using partially quenched heavy baryon chiral
perturbation theory, we derive the chiral and momentum behavior of the
axial and induced pseudoscalar form factors. The meson pole contributions
to the latter posses a striking signal for chiral physics. We argue that
the study of hyperon axial matrix elements enables a systematic lattice
investigation of the efficacy of three flavor chiral expansions in the
baryon sector. This can be achieved by considering chiral corrections to $SU(3)$
symmetry predictions, and their partially quenched generalizations. 
In particular, despite the presence of eight unknown low-energy
constants, we are able to make next-to-leading order symmetry breaking
predictions for two linear combinations of axial charges.
  
\end{abstract}

\maketitle

\section{Introduction}

For the last decade, lattice gauge theory techniques have made 
dramatic progress in increasing our understanding of the
non-perturbative regime of QCD \cite{DeGrand:2007}. 
Despite considerable advances, there are still sources of systematic error in 
lattice data, for example, the finite extent of the lattice and the 
unphysically large quark masses. Fortunately low-energy hadron properties are
dominated by virtual pion interactions and the systematic treatment of such 
interactions using chiral perturbation theory ($\chi$PT) allows one to
parametrize the lattice volume and quark mass dependence of certain observables. 
There has been considerable activity to
understand theoretically the quark mass and lattice volume dependence of 
hadronic observables. 
Further extentions of chiral perturbation theory have been developed to
account for quenching and partially quenching \cite{Bernard:1992,Bernard:1994,
Sharpe:2000,Sharpe:2001fh}, and  
discretization errors \cite{Sharpe:1998,Sharpe:1999}.
An example is the nucleon axial charge, $g_A$. 
Recent lattice studies have made impressive strides toward determining 
$g_A$ \cite{Edwards:2005,Khan:2006}.
In tandem, recent $\chi$PT analyses of the chiral \cite{
s&m:2002,Hemmert:2003,Hemmert:2006}, continuum \cite{s&m:2003,fjj:2007}
and volume extrapolations \cite{s&m:2004,d&m:2004,s&w:2007} are poised to 
connect the data to the physical point.
We are beginning to enter a stage in which the combination of lattice 
QCD data and $\chi$PT will enable the study the 
hadronic properties from the first principles.

A serious issue, however,  confronts this program when extended to hyperon 
observables. Various $SU(3)$ predictions for hyperon properties compare 
poorly to experiment in contrast to the many successful $SU(2)$ 
predictions for the nucleon. While $\chi$PT can be used to systematically 
incorporate effects from the strange quark mass, the systematic expansion 
in the baryon sector has terms that scale (in the worst case) as 
$\sim m_\eta / M$, where $m_\eta$ is the mass of the $\eta$-meson
and $M$ is the average hyperon mass. A well known conflict between 
$\chi$PT analyses and experimental data exist for hyperon decays.
For example, the non-leptonic weak decays, $\Lambda \to p \pi^-$
and $\Sigma^+ \to n \pi^+$, have been extensively investigated 
experimentally. In particular the $s$- and $p$-wave 
contributions to these weak decays are determined to high precision.
Although efforts in the framework of $\chi$PT have been devoted to understand 
these non-leptonic decays theoretically
\cite{Jenkins:1991jv,
Jenkins:1991es,Jenkins:1992,
Springer:1995,m&w:1997,Borasoy:1999.0,Springer:1999.1,AbdElHady:1998gw,
AH:1999,bbps:2005},
long-standing disagreement between these theoretical analyses and 
experimental data remain \cite{Gasp:2003,Hint:2003,Silas:2006.4}.

One is thus led to question the efficacy of three-flavor $\chi$PT
in the baryon sector. Without this systematic model-independent expansion, 
lattice QCD data for hyperon properties cannot be reliably extrapolated 
to the physical values of the quark masses. Additionally volume and continuum 
extrapolations using three-flavor \CPT\ cannot be trusted.
Indeed the first lattice 
calculation of hyperon axial charges, 
$g_{\Sigma\Sigma}$ and $g_{\Xi \Xi}$ \cite{Lin:2007gv,Lin:2007ap},
shows little evidence for the one-loop predictions from (partially quenched) 
$\chi$PT \cite{will:2004}. The lattice, however, can provide a diagnostic tool 
to investigate the condition of three-flavor $\chi$PT.
A complete study of baryon axial charges is the natural starting point.  
These couplings enter in the loop graphs that determine the long-range 
chiral corrections to all baryon observables. Input of these measured 
parameters into $\chi$PT expressions allows one to numerically assess
the behavior of the long-range contributions in the chiral expansion.
This information can then be used to address the convergence of the 
chiral expansion. Perhaps the expansion is converging to the wrong answer, 
or perhaps the expansion is not converging at all. 
If it is the latter case, one can use the lattice to investigate the
cause. Perhaps certain observables are corrupted by 
large values of local contributions that can be isolated
and determined from lattice data, or perhaps nearby resonances 
are leading to large enhancements.

In this work, we provide a follow-up to \cite{will:2004}
by determining the full set of hyperon matrix elements of flavor changing 
axial currents. We work to next-to-leading order in partially 
quenched heavy baryon chiral perturbation theory to address both the chiral
behavior and momentum-transfer dependence of the axial form factors.
Due to meson pole contributions, the pseudoscalar form factor provides 
an observable 
well-suited for the investigation of chiral physics in three-flavor theories. 
Despite the accumulation of a large number (eight) of undetermined 
low-energy constants, we utilize the full set of axial charges to make 
non-trivial next-to-leading order predictions.

Our paper is organized as follows. First in Sect. II, aspects of
PQ$\chi$PT relevant to our calculations are reviewed. 
In Sect. III, we map the PQQCD axial-vector current onto operators in 
PQ$\chi$PT up to next-to-leading order. 
The hyperon axial-current matrix elements are determined for 
$|\D I| = 1$ transitions (Sect. III B), and $|\D S| = 1$ transitions 
(Sect. III C). Various wavefunction renormalization factors
are collected in the Appendix.
Non-trivial next-to-leading order predictions for axial
charges, and a discussion of $SU(3)$
breaking corrections are presented in Sect. IV, which concludes our paper. 

\section{Partially  Quenched Chiral Lagrangian}
\label{chipt}
Before we detail the calculation of the axial current matrix elements, 
we briefly review partially quenched chiral 
perturbation theory. 
We recall the partially quenched chiral Lagrangian in 
the meson sector first and emphasize the relation between lattice measured
meson masses and the parameters of the Lagrangian. The baryon 
Lagrangian is then described in detail. 

\subsection{Mesons}
The lattice action we consider here is comprised of valence and sea 
quarks, each of which comes in three flavors. In the continuum limit,
this action can be described by the partially quenched QCD (PQQCD) 
Lagrange density, which is given by
\begin{equation}
\cL = \ol Q\, i\Dslashe \, Q - \ol Q m_Q Q\, ,
\label{pqqcd}
\end{equation} 
where the quark fields appear in the vector $Q$, which has entries
\begin{equation}
Q = ( u, d, s, j, l, r, \tilde{u}, \tilde{d}, \tilde{s} )^T\,,
\label{3flavor}
\end{equation}
and transforms in the fundamental representation of the graded group
$SU(6|3)$. 
The quark components of the field $Q$ satisfy the following 
graded equal-time commutation relation
\begin{eqnarray}
Q_i^\alpha ({\bf x}) Q_k^{\beta \dagger} ({\bf y}) - 
(-)^{\eta_i\eta_k}Q_k^{\beta \dagger} ({\bf y})Q_i^\alpha ({\bf x})
& = & 
\delta^{\alpha\beta}\delta_{ik}\delta^3({\bf x}-{\bf y})\, ,
\label{eq:comm}
\end{eqnarray}
where ($\alpha$,$\beta$) and ($i$,$k$) are spin and flavor indices 
respectively.
The $\eta_k$'s appearing above are given by
$\eta_k=+1$ for $k=1$--$6$ and $\eta_k=0$ 
for $k=7$--$9$. The $\eta_{k}$ maintain the graded structure of the 
Lie algebra. Further, the graded equal-time 
commutation relations for two $Q$'s or two $Q^\dagger$'s vanish. 
The partially quenched generalization of the mass 
matrix $m_Q$ is given by
\begin{equation}
m_Q = \diag (m_u, m_d, m_s, m_j, m_l ,m_s, m_u, m_d, m_s)\,.
\label{massmatrix1}
\end{equation}
In this work, we enforce the isospin limit in both the valence and sea 
sectors so that we have
\begin{equation}
m_Q = \diag (\bar{m}, \bar{m}, m_s, m_j, m_j ,m_s, \bar{m},\bar{m}, m_s)\,.
\label{massmatrix2}
\end{equation}
Notice that with
Eq.~(\ref{massmatrix1}) (and similarly Eq.~(\ref{massmatrix2})), there is an 
exact cancelation 
between valence and ghost quark contributions to the determinant in the
path integral for the QCD partition function. This cancelation leaves
only the contribution from the sea sector.
When $m_Q = 0$, the Lagrangian Eq.~(\ref{pqqcd})
has a graded $U(6|3)_{L} \otimes U(6|3)_{R}$ symmetry which will reduce to
$SU(6|3)_{L} \otimes SU(6|3)_{R} \otimes U(1)_{V}$ by the axial anomaly
\cite{Sharpe:2001fh}. We assume that the chiral symmetry is 
spontaneously broken:
$SU(6|3)_{L} \otimes SU(6|3)_{R} \rightarrow SU(6|3)_{V}$,
hence an identification between PQQCD and QCD 
can be made.  
The low-energy effective theory of PQQCD is written in 
terms of the pseudo-Goldstone mesons emerging from spontaneous chiral symmetry 
breaking. At leading order in an expansion in momentum and quark mass, 
\footnote{Here we adopt the standard power counting: 
$\partial^2 \sim m_q \sim \varepsilon^2$, where $\varepsilon$ is a small
parameter.} the 
PQ$\chi$PT Lagrangian for the 
mesons is given by
\begin{equation} \label{eq:Llead}
\cL =  
\frac{f^2}{8}
\str \Big(\partial^{\mu}\Sigma^\dagger \partial_\mu\Sigma\Big)
    \,+ \,\lambda\str\Big(m_q\Sigma^\dagger+m_q^\dagger\Sigma\Big)\,-\, 
m_0^2\Phi_0^2\,,
\end{equation}
where $f=132$~MeV, the str() denotes a graded flavor trace and the meson 
fields is incorporated in $\Sigma$ through
\begin{equation}\label{meson}
 \Sigma=\exp\Bigg(\frac{2i\Phi}{f}\Bigg)
  = \xi^2 \, , \ 
  \Phi=
    \Bigg(
      \begin{array}{cc}
        M & \chi^{\dagger} \\ 
        \chi & \tilde{M}
      \end{array}
    \Bigg)
\,.\end{equation}
The matrices $M$, 
$\tilde{M}$ in Eq.~(\ref{meson}) contain bosonic mesons, while $\chi$ and 
$\chi^{\dagger}$ are matrices consisting of fermionic mesons. 
Here $\Phi_{0}\,=\,\str(\Phi)\sqrt{6}$ is the flavor singlet field and is 
included as a device to obtain the flavor neutral propagators in PQ$\chi$PT.
Expanding the Lagrangian in Eq.~\eqref{eq:Llead}, 
one can determine the meson masses which enter into the calculations of baryon 
observables. In particular, the masses of mesons at leading order with 
quark content $Q_{i}\overline{Q}'_{j}$ are
\begin{equation}
m_{Q_{i} Q'_{j}}^2 = \frac{4\lambda}{f^2}((m_{Q})_{ii}\,+\,(m_{Q'})_{jj})\, . 
\label{qmass}
\end{equation}
The flavor singlet field additionally acquires a mass $m^2_{0}$.
Due to the strong $U(1)_{A}$ anomaly, this mass can be taken on the 
order of the chiral symmetry breaking scale, $m_{0}\sim \Lambda_{\chi}\approx
4\pi f$. The flavor singlet field can thus be integrated out. 
However, the propagator of the flavor neutral fields deviate 
from a simple pole form \cite{Sharpe:2001fh}. For 
$a,b = u,d,s$, the $\eta_{a} \eta_{b}$ propagator at leading order is given by 
\cite{Sharpe:2001fh}
\begin{eqnarray}
{\cal G}_{\eta_a\eta_b} & = & 
{ i\delta^{ab}\over q^2-m_{aa}^2+i\epsilon}
\ -\ {i\over 3}
{(q^2-m_{jj}^2+i\epsilon)(q^2-m_{rr}^2+i\epsilon)\over (q^2-m_{aa}^2+i\epsilon)(q^2-m_{bb}^2+i\epsilon)(q^2-m^2_{X}+i\epsilon)}
\,,
\label{doublepole}
\end{eqnarray}
where the masses of valence-valence mesons $m_{aa}^2$, $ m_{bb}^2$ and the
masses of the sea-sea mesons $m_{jj}^2$, $m_{rr}^2$ are  
given by Eq.~(\ref{qmass}). In Eq.~(\ref{doublepole}), the mass $m_X$ is 
defined as $m^2_{X} = \frac{1}{3}(m^2_{jj}+2m^2_{rr})$. The flavor 
neutral propagator Eq.~(\ref{doublepole}) can be conveniently written in 
the form
\begin{eqnarray}
{\cal G}_{\eta_a\eta_b} & = &  \delta^{ab}  P_a\ +\ 
{\cal H}_{aa}( P_a , P_b , P_X)
\,,
\label{doublecomp}
\end{eqnarray}
with
\begin{eqnarray}
P_a & = & { i \over q^2 - m_{aa}^2+i\epsilon}
\, \, ,\, \, 
P_b \, = \,  { i \over q^2 - m_{bb}^2+i\epsilon}
\, \, ,\,\, 
P_X \, = \,  { i \over q^2 - m_X^2+i\epsilon}\,\,,
\nonumber\\
{\cal H}_{ab}( A, B, C) & = & 
-{1\over 3}\left[\ 
{(m_{jj}^2-m_{aa}^2)(m_{rr}^2-m_{aa}^2)\over 
(m_{aa}^2-m_{bb}^2)(m_{aa}^2-m_X^2)}\  A
-
{(m_{jj}^2-m_{bb}^2)(m_{rr}^2-m_{bb}^2)\over 
(m_{aa}^2-m_{bb}^2)(m_{bb}^2-m_X^2)}\  B
\right.\nonumber\\ & & \left.\qquad
\ +\ 
{(m_X^2-m_{jj}^2)(m_X^2-m_{rr}^2)\over 
(m_X^2-m_{aa}^2)(m_X^2-m_{bb}^2)}\  C
\ \right]
\,.
\label{doublecomp1}
\end{eqnarray}                        
The above form is convenient for contributions from flavor neutral mixing.
When there is no mixing, i.e. $a \,=\, b$ and $A \,=\, B$, 
the hairpin propagator
has a double pole and the limit of Eq.~(\ref{doublecomp1}) must be taken, 
and produces 
\begin{eqnarray}
{\cal H}_{aa}(A,A,C) & = &  
-\frac{1}{3} \Bigg[\, \frac{\partial}{\partial {m^{2}_{aa}}} \frac{(m^{2}_{jj}-m^{2}_{aa})(m^{2}_{rr}-m^{2}_{aa})}{(m^{2}_{aa}-m^{2}_{X})} A \nonumber \\
&& \qquad \qquad \quad +  \frac{(m^{2}_{jj}-m^{2}_{X})(m^{2}_{rr}-m^{2}_{X})}{(m^{2}_{X}-m^{2}_{aa})^{2}} C\,\Bigg] \ \ .
\label{doubleisospin}
\end{eqnarray}

In partially quenched simulations, one numerically determines
the values of the valence pion $m_{\pi,\text{val}}$ and valence kaon
$m_{K,\text{val}}$ masses, as well as the sea pion $m_{\pi,\text{sea}}$
and sea kaon $m_{K,\text{sea}}$ masses. When one uses PQ$\chi$PT to calculate 
the meson mass dependence of observables, they are expressed in
terms of meson masses via the tree-level relation in Eq.~(\ref{qmass}).  
To use the lattice determined meson masses in the valence and sea sectors,
it is straightforward algebra to convert the loop meson masses appearing 
in PQ$\chi$PT to those measured directly on the lattice.
Explicitly we have
\begin{eqnarray}
m_{uu}^2
&=&
m_{\eta_u}^2
=
m_{\pi,\text{val}}^2\,,
\nonumber \\
m_{us}^2
&=&
m_{K,\text{val}}^2\,,
\nonumber \\
m_{ss}^2
&=&
 m_{\eta_s}^2
=
2 m_{K,\text{val}}^2 - m_{\pi,\text{val}}^2\,,
\nonumber \\
m_{uj}^2
&=&
\frac{1}{2}
\left(
m_{\pi,\text{val}}^2
+
m_{\pi,\text{sea}}^2
\right)\,,
\nonumber \\
m_{ur}^2
&=&
\frac{1}{2}
\left(
m_{\pi,\text{val}}^2
-
m_{\pi,\text{sea}}^2
\right)
+
m_{K,\text{sea}}^2\,,
\nonumber \\
m_{sj}^2
&=&\frac{1}{2}
\left(
m_{\pi,\text{sea}}^2
-
m_{\pi,\text{val}}^2
\right)
+
m_{K,\text{val}}^2\,,
\nonumber \\
m_{sr}^2
&=&
- \frac{1}{2}
\left(
m_{\pi,\text{val}}^2
+
m_{\pi,\text{sea}}^2
\right)
+
m_{K,\text{val}}^2
+
m_{K,\text{sea}}^2\,,
\nonumber \\
m_{jj}^2
&=&
m_{\pi,\text{sea}}^2\,,
\nonumber \\
m_{rr}^2
&=&
2 m_{K,\text{sea}}^2
-
m_{\pi,\text{sea}}^2\,,
\nonumber \\
m_X^2
&=&
\frac{4}{3} m_{K,\text{sea}}^2
-
\frac{1}{3} m_{\pi,\text{sea}}^2
\,.
\label{convertlattice}
\end{eqnarray}
These relations must be modified if the source of partial quenching is due
to mixed lattice actions, see \cite{Bar:2005tu,Tiburzi:2005is,Chen:2007ug}.
\subsection{Baryon}
In this section, we discuss the baryon sector of PQ$\chi$PT in the framework
of \cite{Labrenz:1996jy,MSavage:2002,MSavage:2002.1}.  
Building blocks for the baryon Lagrangian are the 
super-multiplets $\cB_{ijk}$ and $\cT^{\mu}_{ijk}$. The 
$\bf{240}$-dimensional super-multiplet of spin-$\frac{1}{2}$ baryons 
$\cB_{ijk}$ satisfies the following 
relations under the interchange of the flavor indices \cite{Labrenz:1996jy}
\begin{eqnarray}
{\cal B}_{ijk} & = & (-)^{1+\eta_j \eta_k}\,  {\cal B}_{ikj}
\, , \,
{\cal B}_{ijk} \, +\,  (-)^{1+\eta_i \eta_j}\, {\cal B}_{jik}
\, +\, (-)^{1 + \eta_i\eta_j + \eta_j\eta_k + \eta_k\eta_i}\, 
{\cal B}_{kji}\, =\, 0\,,
\label{eq:bianchi1}
\end{eqnarray}
and the familiar octet baryons are embeded in $\cB_{ijk}$ through
\cite{MSavage:2002,MSavage:2002.1}
\begin{equation}
{\cal B}_{ijk}  = \frac{1}{\sqrt{6}}(\epsilon_{ijl}B_{k}^{l}  +
\epsilon_{ikl}B_{j}^{l})\,,
\end{equation} 
where $B$ is the octet baryon matrix
\begin{equation}
   B = 
    \begin{pmatrix}
      \frac{1}{\sqrt{6}}\Lambda + \frac{1}{\sqrt{2}}\Sigma^0 &
        \Sigma^+ & p\\
      \Sigma^- & \frac{1}{\sqrt{6}}\Lambda -
        \frac{1}{\sqrt{2}}\Sigma^0 & n\\
      \Xi^- & \Xi^0 & -\frac{2}{\sqrt{6}}\Lambda\\
    \end{pmatrix} \,.
\label{eq:octetbaryons}
\end{equation}
The spin-$\frac{3}{2}$ resonances are contained in the
the $\bf{138}$-dimensional super-multiplet $\cT_{ijk}$, which
satisfies 
\begin{eqnarray}
{\cal T}_{ijk} \, = \,  
(-)^{1+\eta_i\eta_j} {\cal T}_{jik}\, =\, 
(-)^{1+\eta_j\eta_k} {\cal T}_{ikj} \,,
\label{eq:bianchi2}
\end{eqnarray}
under the interchange of flavor indices \cite{Labrenz:1996jy}. 
Furthermore, one embeds the decuplet baryons in $\cT_{ijk}$ by  
\begin{equation}
\ \ \ \ {\cal T}_{ijk} = T_{ijk} \,, 
\end{equation}
where $T$ is totally symmetric tensor containing the decuplet resonances
\begin{eqnarray}
T_{111} \,& = &\, \Delta^{++}
\,\, ,\,\,
T_{112} \, =\,  {1\over\sqrt{3}}\Delta^+
\,\, ,\,\,
T_{122} \, =\,  {1\over\sqrt{3}}\Delta^0
\,\, , \,\, 
T_{222} \, =\,   \Delta^{-}\,,
\nonumber\\ 
T_{113} \,& = &\, {1\over\sqrt{3}}\Sigma^{*,+}
\,\, ,\,\, 
T_{123} \, =\,  {1\over\sqrt{6}}\Sigma^{*,0}
\,\, ,\,\, 
T_{223} \, =\,  {1\over\sqrt{3}}\Sigma^{*,-}\,,
\nonumber\\
T_{133} \,& = &\, {1\over\sqrt{3}}\Xi^{*,0}
\,\, ,\,\, 
T_{233} \, =\,  {1\over\sqrt{3}}\Xi^{*,-}
\,\, , \,\, 
T_{333} \, =\,  \Omega^{-}
\, .
\label{eq:decupletbaryon}
\end{eqnarray}
The free Lagrangian for the $\bf{240}$-dimensional super-multiplet $\cB_{ijk}$ 
and the $\bf{138}$-dimensional super-multiplet $\cT_{ijk}$ fields in 
$SU(6|3)$ PQ$\chi$PT is given by 
\cite{MSavage:2002.1}
\begin{eqnarray} \label{eqn:L}
  {\mathcal L}
  &=&
   i\Big(\ol\cB v\cdot{\mathcal D}\cB\Big)
  +2\a_{M}\Big(\ol\cB \cB{\mathcal M}_+\Big)
  +2\b_{M}\Big(\ol\cB {\mathcal M}_+\cB\Big)
  +2\sigma_{M}\Big(\ol\cB\cB\Big)\str\Big({\mathcal M}_+\Big)  \nonumber \\
  &-&i\Big(\ol\cT^{\mu} v\cdot{\mathcal D}\cT_\mu\Big)
    +\D\Big(\ol\cT^{\mu}\cT_\mu\Big)
    +2\g_{M}\Big(\ol\cT^{\mu} {\mathcal M}_+\cT_\mu\Big)
    -2\ol\sigma_{M}\Big(\ol\cT^{\mu}\cT_\mu\Big)\str\Big({\mathcal M}_+\Big)
  \,,
\end{eqnarray}
where the mass operator ${\mathcal M}_{+}$ is defined by:
\begin{equation}
{\mathcal M}_+ = \frac{1}{2}\left(\xi^\dagger m_Q \xi^\dagger + \xi m_Q \xi\right)
\, .\end{equation}
The parameter $\Delta$ is the mass splitting between the octet 
and decuplet baryons in the chiral limt.
Phenomenologically we know  $\Delta \sim m_{\phi}$, where $\phi$ is an $SU(3)$
meson, hence the decuplet baryons much be included as dynamical fields in
Eq.~(\ref{eqn:L}).
The parenthesis notation for flavor contractions used in Eq.~\eqref{eqn:L} 
is that of~\cite{MSavage:2002.1}.
The partially quenched Lagrangian describing the interactions of the 
$\cB_{ijk}$ and $\cT^{\mu}_{ijk}$ with the pseudo-Goldstone mesons is given by
\cite{MSavage:2002.1}
\begin{eqnarray} \label{eqn:Linteract}
  {\cal L} &=&   
	  2 \a \Big(\ol \cB S^{\mu} \cB A_\mu \Big)
	+ 2 \b \Big(\ol \cB S^{\mu} A_\mu \cB \Big)
	+ 2{\mathcal H}\Big(\ol{\cT}^{\nu} S^{\mu} A_\mu \cT_\nu\Big) \nonumber \\ 
    	&+& \sqrt{\frac{3}{2}}\cC
  		\Big[
    			\Big(\ol{\cT}^{\nu} A_\nu \cB\Big)+ \Big(\ol \cB A_\nu \cT^{\nu}\Big)
  		\Big] \,.   
\label{interL}
\end{eqnarray}
The axial-vector and vector meson fields $A_\mu$ and $V_\mu$
are defined by: $ A_{\mu}=\frac{i}{2}
\left(\xi\partial_{\mu}\xi^\dagger-\xi^\dagger\partial_{\mu}\xi\right)$  
and $V_{\mu}=\frac{1}{2} \left(\xi\partial_{\mu}\xi^\dagger+\xi^\dagger\partial_{\mu}\xi\right)$.
The latter appears in  Eq.~\eqref{eqn:L} for the
covariant derivatives of $\cB_{ijk}$ and $\cT_{ijk}$ 
that both have the form
\begin{equation}
  ({\mathcal D}^{\mu} \cB)_{ijk}
  =
  \partial^{\mu} \cB_{ijk}
  +(V^{\mu})^{l}_{i}\cB_{ljk}
  +(-)^{\eta_i(\eta_j+\eta_m)}(V^\mu)^{m}_{j}\cB_{imk}
  +(-)^{(\eta_i+\eta_j)(\eta_k+\eta_n)}(V^{\mu})^{n}_{k}\cB_{ijn}\, .
\end{equation}
The vector $S_{\mu}$ is the covariant spin operator \cite{Jenkins:1991jv,
Jenkins:1991es}. The parameters that appear in the \PQCPT\ Lagrangian can be 
related to those in \CPT\ by matching. To be more specific, one restricts to 
the $q_{\text{sea}}q_{\text{sea}}q_{\text{sea}}$ sector and compares the
\PQCPT\ Lagrangian obtained with that  
of $\chi$PT. With this matching procedure, one finds that
$\alpha = \frac{2}{3}D + 2F$, $\beta = -\frac{5}{3}D + F$, and the other 
parameters ${\cal C}$ and ${\cal H}$ appearing above have the same 
numerical values as in $\chi$PT \cite{MSavage:2002.1}.

\section{The Axial-Vector Current} 
\label{accurent}
\subsection{The Axial-Vector Current in PQ$\chi$PT}
\label{accurentpqcpt}
The baryon matrix elements of the axial-vector current, 
$j^{a}_{\mu,5} = \overline{q}\lambda^{a}\gamma_\mu\gamma_5 q$, have been studied 
extensively both 
on the lattice \cite{Edwards:2005,Khan:2006} and 
$\chi$PT \cite{Jenkins:1991jv,Jenkins:1991es,Kim&Kim:1996,
Borasoy:1999,Zhu:2001,s&m:2002,s&m:2004,dwill:2002,d&m:2004,will:2004,
Kambor:1999,Bernard:1998gv, Hemmert:2003,Meissner:2006,Hemmert:2006}. 
In PQQCD, the axial
current is defined by $J^{a}_{\mu,5} = 
\overline{Q}\,\overline{\lambda}{}^{a}\gamma_\mu\gamma_5 Q$. 
In general, one must worry that the choice of supermatrices
$\overline{\lambda}{}^{a}$ is not unique even after the 
requirement  $\text{str}(\overline{\lambda}{}^{a}) = 0$ 
has been enforced. To be relevant for any practical lattice calculation, the
choice of PQQCD matrices should maintain the cancelation of valence and ghost
quark loops with an operator insertion \cite{will:2004,bct0412}. This is
because otherwise the PQQCD theory corresponds to a lattice theory where twice
the number of disconnected contractions must be calculated. However 
since we are only interested in flavor-changing operators, the self 
contractions of these
operators automatically vanish. Thus we can decouple the ghost and sea quarks
sectors from the flavor changing axial current by choosing the upper
$3\times3$ block of $\overline{\lambda}$ to be the Gell-Mann matrices. This
choice merely corresponds to an axial transition operator that only acts in
the valence sector and is precisely what is implemented on the
lattice.\footnote{Isospin symmetry allows one to relate isospin 
transition matrix elements to differences of flavor conserving matrix elements.
These difference have often been calculated  on the lattice.
For strangeness transitions, $SU(3)$ is badly broken disallowing
the analogous procedure.}

Having fixed the $\overline{\lambda}$ supermatrices, we map the PQQCD axial
current operator into the heavy baryon PQ$\chi$PT. At leading order, the 
PQ$\chi$PT axial current is given by \cite{s&m:2002}
\begin{eqnarray}
J_{\mu,5}^{a}
& = & 
2\alpha\Big(\overline{\cal B} S_\mu {\cal B}{\overline{\lambda}{}^{a}_{\xi +}}
\Big)\,+\, 
2\beta\Big(\overline{\cal B} S_\mu {\overline{\lambda}{}^{a}_{\xi +}}{\cal B}
\Big)\,+\,  
2{\cal H}\Big(\overline{\cal T}^\nu S_\mu {\overline{\lambda}{}^{a}_{\xi +}}{\cal T}_\nu\Big)\nonumber\\
&\,+\,&
\sqrt{3\over 2}{\cal C} 
\Big[\Big(\overline{\cal T}_\mu {\overline{\lambda}{}^{a}_{\xi +}} {\cal B}\Big)
\,+\,\Big(\overline{\cal B}\,{\overline{\lambda}{}^{a}_{\xi +}} {\cal T}_\mu\Big)
\Big].
\label{eq:LOaxialcurrent}
\end{eqnarray}
with $\alpha$, $\beta$, ${\cal H}$ and ${\cal C}$ the same low energy constants
(LEC's) appearing in Eq.~(\ref{eqn:Linteract}) and 
$\overline{\lambda}{}^a_{\xi +} = {1\over 2}(
\xi\overline{\lambda}{}^a\xi^\dagger
+\xi^\dagger\overline{\lambda}{}^a\xi)$. 
Since we work to next-to-leading order (NLO) in the chiral expansion
and NLO in the momentum expansion, 
we further require the contributions to the matrix elements from NLO
axial current. 
At NLO, there are two contributions to the axial matrix 
elements: one is from the NLO axial current in the baryon sector and the other
is obtained from the local counterterms involving one insertion of the quark 
mass matrix $m_{Q}$.
The former is given by
\begin{eqnarray}
J^{a}_{\mu,5} 
&=&
\frac{1}{\L_\chi^2} 
\Bigg\{ 
2 n_\a 
\Big[
\partial_{\mu} \partial_{\nu} 
\Big( \overline{\cB} S^\nu \cB \, \overline{\lambda}^{a}_{\xi^{+}} \Big)
-
\partial^2
\Big( \overline{\cB} S_\mu \cB \, \overline{\lambda}^{a}_{\xi^{+}}  \Big)
\Big] \nonumber \\
& + &
2 n_\b 
\Big[
\partial_{\mu} \partial_{\nu} 
\Big( \overline{\cB} S^\nu \, \overline{\lambda}^{a}_{\xi^{+}} \cB \Big)
-
\partial^2
\Big( \overline{\cB} S_\mu \, \overline{\lambda}^{a}_{\xi^{+}} \cB \Big)
\Big]
\Bigg\}\,.
\label{nloPQ}
\end{eqnarray}
while the latter reads \cite{s&m:2002}:
\begin{eqnarray}
J_{\mu,5}^{a,m_{Q}}&=& 
16\frac{\lambda}{f^2}\Big[ 
b_1 \cbb^{kji}\{ \overline{\lambda}^a_{\xi +}\,,\, 
{\cal M}_+\}^n_i\ S_\mu \cb_{njk}
+b_2 (-)^{(\eta_i+\eta_j)(\eta_k+\eta_n)} 
\cbb^{kji}\{\overline{\lambda}^a_{\xi +}\,,\, {\cal M}_+\}^n_k 
 S_\mu \cb_{ijn}\nonumber\\
&+& b_3(-)^{\eta_l (\eta_j+\eta_n)}
\cbb^{kji} (\overline{\lambda}^a_{\xi +})^l_i 
({\cal M}_+)^n_j S_\mu \cb_{lnk} \nonumber\\
&+& b_4(-)^{\eta_l \eta_j + 1} 
\cbb^{kji} \Big(  
(\overline{\lambda}^a_{\xi +})^l_i ( {\cal M}_+)^n_j
+( {\cal M}_+)^l_i (\overline{\lambda}^a)^n_j \Big)S_\mu \cb_{nlk}
\nonumber\\
&+&b_5 (-)^{\eta_i(\eta_l+\eta_j)}
\cbb^{kji}(\overline{\lambda}^a_{\xi +})^l_j 
( {\cal M}_+)^n_i S_\mu \cb_{nlk}
+b_6 \cbb^{kji} (\overline{\lambda}^a_{\xi +})^l_i S_\mu \cb_{ljk}
\ {\rm str}( {\cal M}_+) \nonumber\\
&+& b_7(-)^{(\eta_i+\eta_j)(\eta_k+\eta_n)}
\cbb^{kji} (\overline{\lambda}^a_{\xi +})^n_k S_\mu \cb_{ijn}
\ {\rm str}( {\cal M}_+) 
\nonumber\\
&+& b_8 \cbb^{kji} S_\mu \cb_{ijk} 
\ {\rm str}(\overline{\lambda}^a_{\xi +} {\cal M}_+ ) \Big]\,,
\label{eq:axcts}
\end{eqnarray}
where the coefficients $b_1,b_2.....,b_8$ must be determined from lattice
simulations. The relation between the partially quenched parameters 
$n_{\alpha}\,,n_{\beta}$ 
and the physical parameters $n_{D}\,,n_{F}$ in usual $SU(3)$ $\chi$PT can be 
obtained by matching: 
$n_{\alpha} = \frac{2}{3}n_{D} + 2n_{F}$, 
$n_{\beta} = -\frac{5}{3}n_{D} + n_{F}$.
Notice that the operator $b_8 \cbb^{kji} S_\mu \cb_{ijk} 
\ {\rm str}(\overline{\lambda}^a_{\xi +} {\cal M}_+ )$ does not contribute 
to the flavor changing transitions at
tree-level. This leaves seven independent partially quenched NLO operators, 
one more than that in ordinary $SU(3)$. However, because these counterterms only
contribute to tree-level, no unphysical combinations will be introduced.
\begin{figure}
\begin{center}
\hbox{
\includegraphics[width=0.4\textwidth]{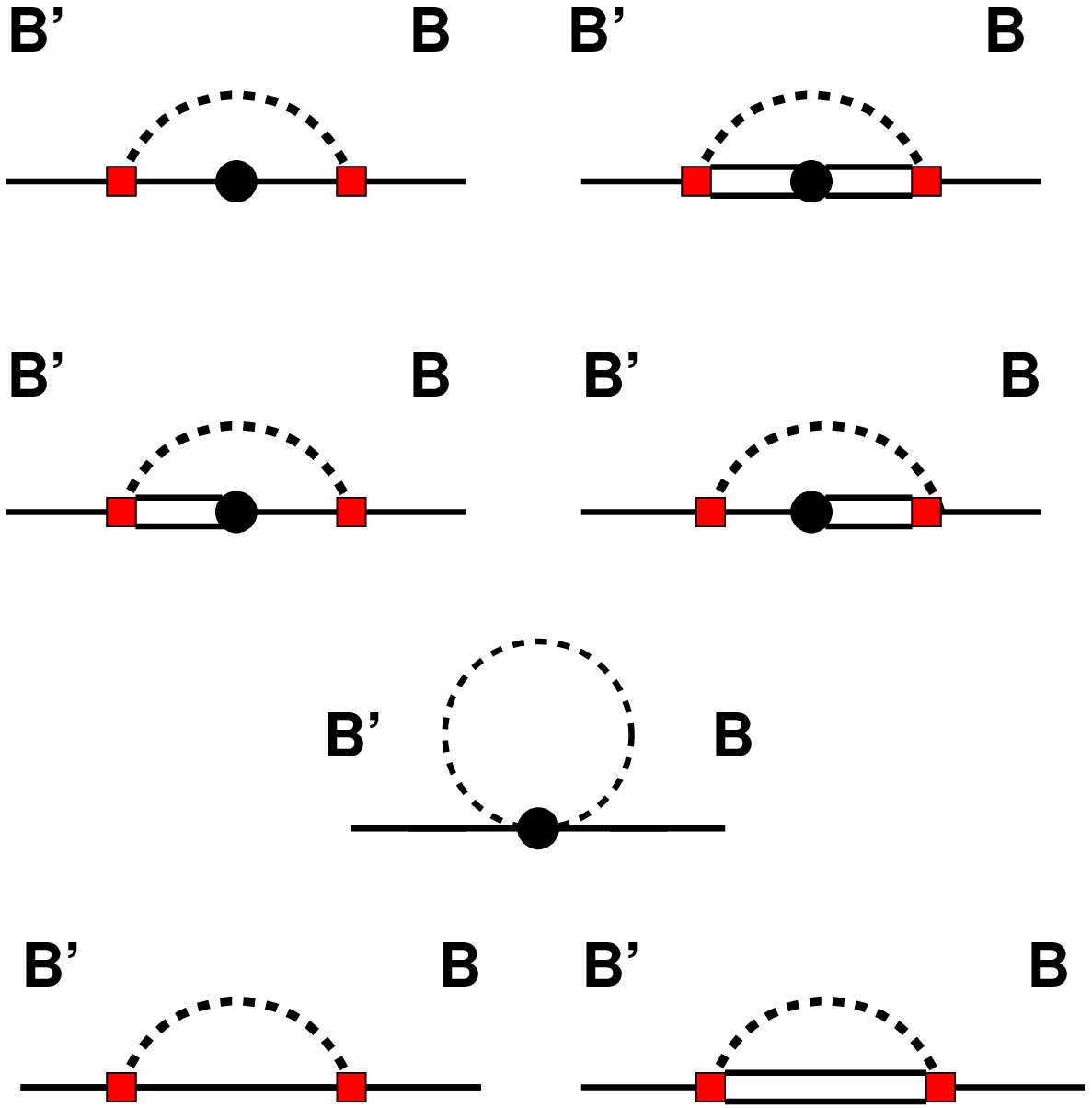}
~~~~~~~~
\includegraphics[width=0.4\textwidth]{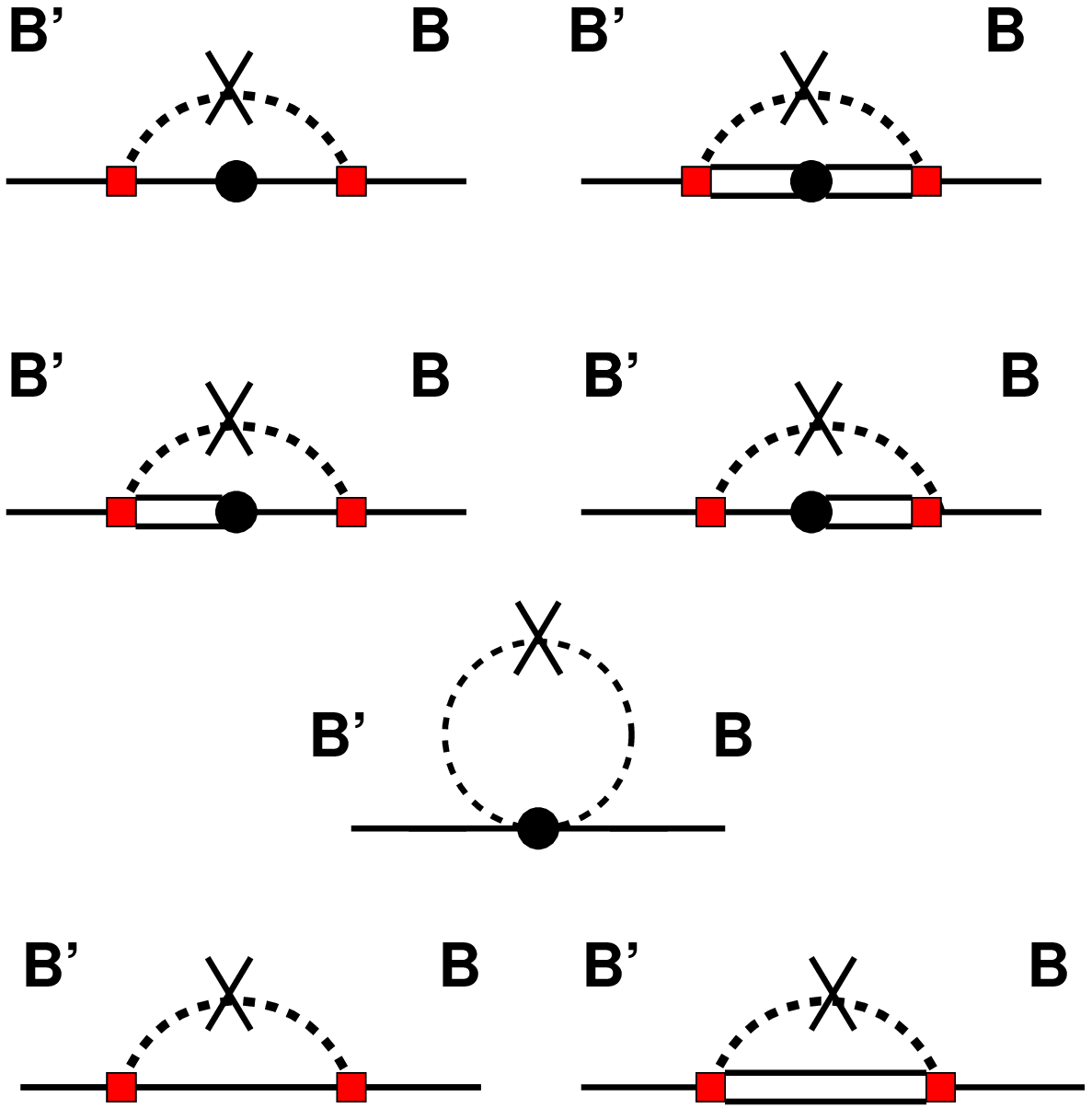}
}
\end{center}
\caption{One-loop diagrams which contribute to 
the leading non-analytic terms of the octet baryon axial form factors. 
Mesons are represented by a dashed line while the single 
and double lines are the symbols for an octet and a decuplet respectively. 
The solid circle is an insertion of the axial current operator and the
solid squares are the couplings given in Eq.~(\ref{interL}). 
The wave function renormalization diagrams are depicted 
in the bottom row. The diagrams with a cross on the loop meson are 
the hairpin contributions which arise from the flavor neutral meson 
propagators.}
\label{fig0}
\end{figure}

\begin{figure}
\begin{center}
\hbox{
\includegraphics[width=0.4825\textwidth]{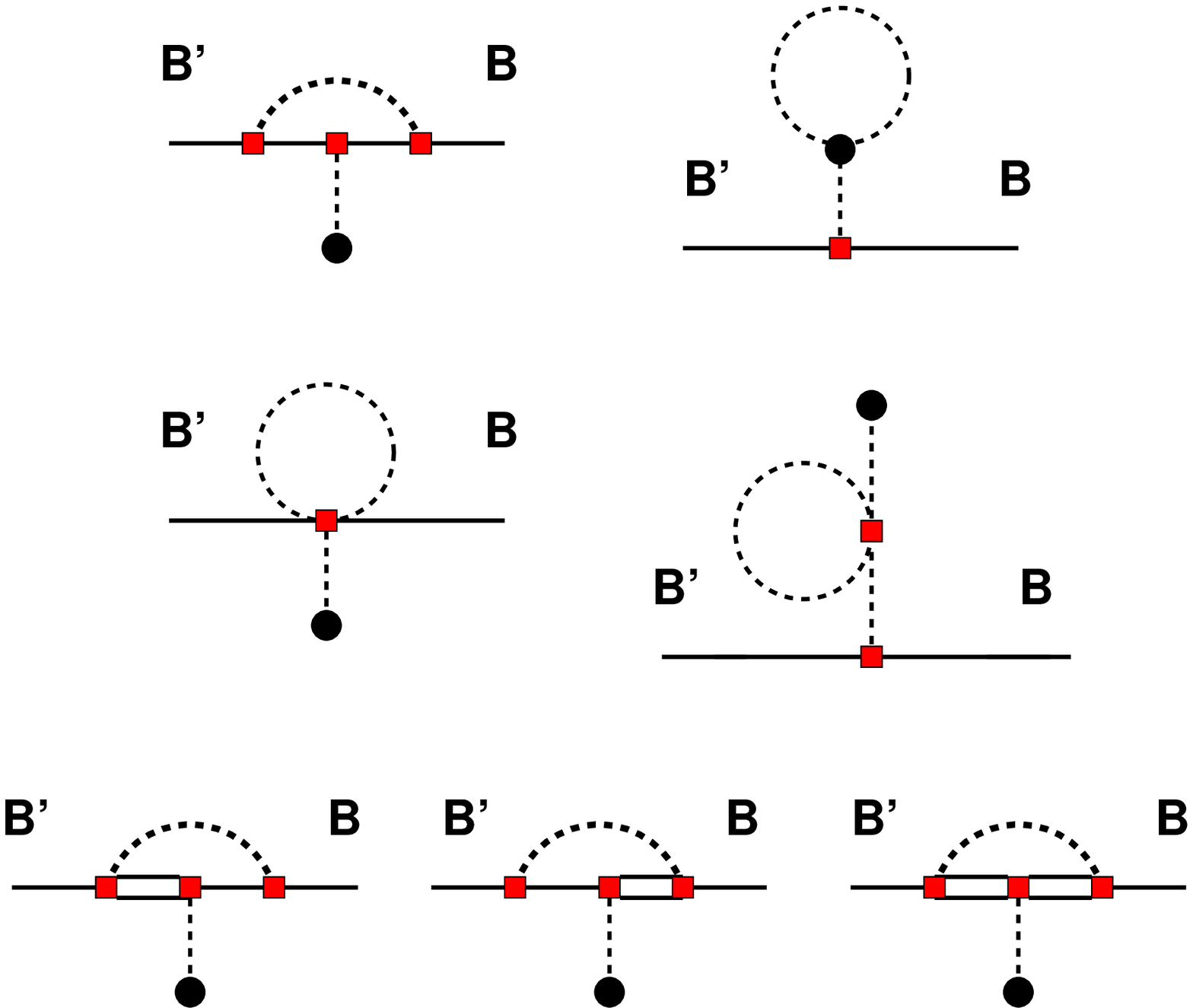}
~~~~
\includegraphics[width=0.4825\textwidth]{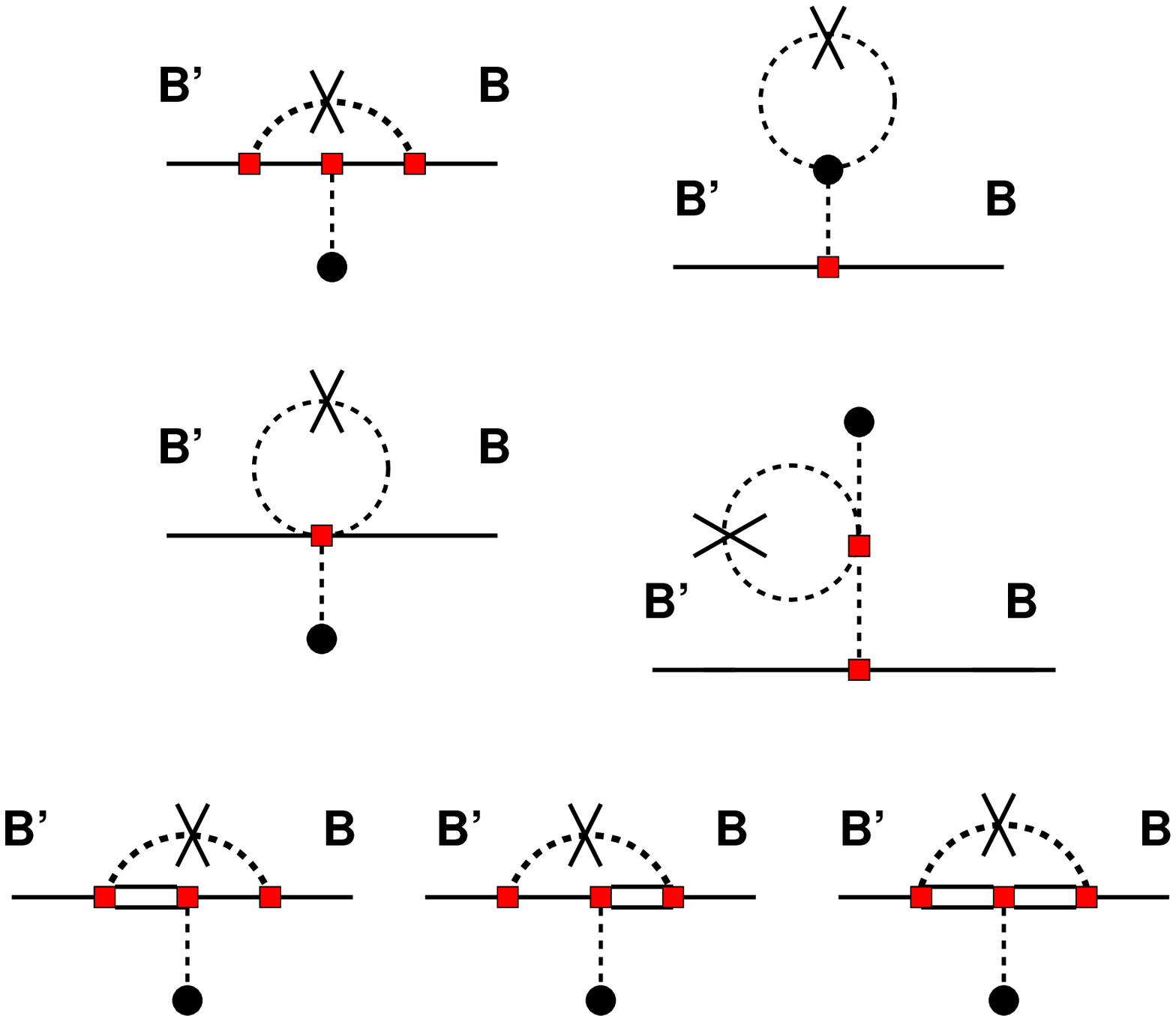}
}
\end{center}
\caption{One-loop diagrams which contribute to 
the leading non-analytic terms of the octet baryon induced pseudoscalar 
form factors. Diagram elements are the same as FIG.~\ref{fig0}} 
\label{fig1}
\end{figure}

\subsection{Isospin Changing Transitions}
\label{IST}
In the partially quenched theory, the supermatrix for the $\Delta I \,=\,1$ 
isospin changing transitions is
\begin{equation} \label{eqn:tau}
\overline{\lambda}^{1+2i}_{i j} = 
\begin{cases}
 1 \quad i = 1\, , j = 2 \\
 0  \quad {\text{otherwise}}
\end{cases}
\,.\end{equation}
Within the baryon octet there are six isospin $\Delta I = 1$ changing 
transitions, namely
\begin{equation}
n \rightarrow p\,\,,\,\,\Sigma^{0} \rightarrow \Sigma^{+}\,\,,\,\,
\Sigma^{-} \rightarrow \Sigma^{0}\,\,,\,\,\Lambda 
\rightarrow \Sigma^{+}\,\,,\,\,
\Sigma^{-} \rightarrow \Lambda\,\,,\,\,\Xi^{-} \rightarrow \Xi^{0}\,\,.
\end{equation}
The neutron to proton axial transition defines the nucleon axial form factor
$G_{A,NN}(q^2)$
and induced pseudoscalar form factor $G_{P,NN}(q^2)$ 
\begin{equation}
\langle p (P') | J^{1+2i}_{\mu,5} | n
(P)\rangle\,=\,\overline{U}_{N}(P')
\Big[2S_{\mu}G_{A,NN}(q^2)+\frac{q_{\mu} q \cdot S}{(2m_{N})^2}G_{P,NN}(q^2) \Big]U_{N}(P)\,,
\end{equation}
where $q_{\mu} = (P' - P)_{\mu}$ is the four momentum transfer. 
The $\Sigma$$\Sigma$ transition matrix elements define the $\Sigma$$\Sigma$ 
axial form factor $G_{A,\Sigma\Sigma}(q^2)$ and induced pseudoscalar form 
factor $G_{P,\Sigma\Sigma}(q^2)$
\begin{equation}
\langle \Sigma^{0} (P') | J^{1+2i}_{\mu,5} | \Sigma^{-}
(P)\rangle\,=\,\frac{1}{\sqrt{2}}\overline{U}_{\Sigma}(P')
\Big[2S_{\mu}G_{A,\Sigma\Sigma}(q^2)+\frac{q_{\mu} q \cdot S}{(2m_{\Sigma})^2}G_{P,\Sigma\Sigma}(q^2) \Big]U_{\Sigma}(P)\,.
\end{equation}
While there are two $\Sigma$$\Sigma$ isospin transitions, their matrix 
elements are related by isospin algebra
\begin{equation}
\langle \Sigma^{+}(P') | J^{1+2i}_{\mu,5} | 
\Sigma^{0}(P)\rangle\,=\,
-\langle \Sigma^{0} (P')| J^{1+2i}_{\mu,5} | \Sigma^{-}(P)\rangle\,.
\end{equation}
The $\Lambda$$\Sigma$ transition matrix elements define the 
$\Lambda$$\Sigma$ axial form factor $G_{A,\Lambda\Sigma}$
and induced pseudoscalar form factor $G_{P,\Lambda\Sigma}(q^2)$
\begin{equation}
\langle \Lambda (P') | J^{1+2i}_{\mu,5} | \Sigma^{-}
(P)\rangle\,=\,\frac{1}{\sqrt{6}}\overline{U}_{\Lambda}(P')
\Big[2S_{\mu}G_{A,\Lambda\Sigma}(q^2)+\frac{q_{\mu} q \cdot S}{(m_{\Lambda}+m_{\Sigma})^2}G_{P,\Lambda\Sigma}(q^2) \Big]U_{\Sigma}(P)\,.
\end{equation}
Although there are two $\Lambda$$\Sigma$ transition matrix elements, they
are related by isospin
\begin{equation}
\langle \Sigma^{+}(P') | J^{1+2i}_{\mu,5} | \Lambda(P)\rangle\,=\,
\langle \Lambda (P')| J^{1+2i}_{\mu,5} | \Sigma^{-}(P)\rangle\,.
\end{equation}
Finally, the $\Xi$$\Xi$ axial form factor $G_{A,\Xi\Xi}$ and induced
pseudoscalar form factor $G_{P,\Lambda\Sigma}(q^2)$
appear in the $\Xi$$\Xi$  transition matrix element
\begin{equation}
\langle \Xi^{0} (P') | J^{1+2i}_{\mu,5} | \Xi^{-}
(P)\rangle\,=\,\overline{U}_{\Xi}(P')
\Big[2S_{\mu}G_{A,\Xi\Xi}(q^2)+\frac{q_{\mu} q \cdot S}{(2m_{\Xi})^2}G_{P,\Xi\Xi}
(q^2) \Big]U_{\Xi}(P)\,.
\end{equation}
Here we use heavy baryon spinors and notation. One can easily show 
up to recoil corrections,
$2\overline{U}(P') S_{\mu} U (P)= \overline{U}(P')\gamma_{\mu} \gamma_{5}
U(P)$, where on the right-hand side appears ordinary Dirac matrices and 
spinors. Thus the axial charges, $G_{A,B'B}(0)$, are the 
standard ones. In our power counting, while the tree-level contributions 
from LO axial current is of order $\varepsilon^{0}$, the tree-level 
contributions obtained from NLO current count $\varepsilon^2$. In addition, 
at order $\varepsilon^2$
there are leading non-analytic contributions to the matrix elements from the
one-loop diagrams shown in Fig.~\ref{fig0} and Fig.~\ref{fig1}. 
The one-loop diagrams in Fig.~\ref{fig0} contribute to the axial form factors
while the induced pseudoscalar form factors receive contributions
from the one-loop diagrams in Fig.~\ref{fig1}. 
Evaluation of the diagrams in Fig.~\ref{fig0} together with the tree-level 
contributions yields the following expression for the axial form factor 
of $\Delta I \,=\,1$ isospin transitions
\begin{eqnarray}
G_{A,B'B}(q^2) & = &\, g_{B'B}\sqrt{Z_{B'}Z_{B}} + \frac{1}{16\pi^2f^2}
\Bigg[ g_{B'B} \sum_{\phi}C_{\phi}{\cal L}(m_{\phi},\mu)
\nonumber \\
& + & {\cal H}{\cal C}^2\Big(\sum_{\phi} E_{\phi}{\cal J}(m_{\phi},\Delta,\mu)
 + \sum_{\phi\phi^{\prime}} \bar{E}_{\phi\phi^{\prime}}{\cal T}(\eta_{\phi}
  \eta_{\phi'},\Delta,\mu)\Big)\nonumber \\
& + & 
\cC^2 
\Big(\sum_{\phi}A_{\phi}{\cal K}(m_{\phi},\Delta,\mu)
 + \sum_{\phi\phi^{\prime}} \bar{A}_{\phi\phi^{\prime}}
{\cal S}(\eta_{\phi} \eta_{\phi^{\prime}},\Delta,\mu)
\Big) \nonumber \\
& + &\sum_{\phi} Y_{\phi}{\cal L}(m_{\phi},\mu)
 + \sum_{\phi\phi^{\prime}} \bar{Y}_{\phi,\phi^{\prime}}{\cal
  R}(\eta_{\phi}\eta_{\phi'},\Delta,\mu) \Bigg] \nonumber \\ 
& + & n_{B'B}\frac{q^2}{\Lambda_{\chi}^2} + \sum_{\phi}u_{\phi}m_{\phi}^2
\,.  
\label{matrixiso}
\end{eqnarray}
In Eq.~(\ref{matrixiso}), $B$ ($B'$) stands for
the initial (final) octet baryon, $Z_{B}$ and $Z_{B'}$ are 
the wave function renormalization factors and are given in the
Appendix. The constants $g_{B'B}$'s are the leading order octet 
baryon axial charges
\begin{eqnarray}
g_{NN} &=& (D\,+\,F)\,, \,\, g_{\Lambda\Sigma} = 2D\,,  
\,\, g_{\Xi\Xi} = (D\,-\,F)\,,\,\,
g_{\Sigma\Sigma} = 2F\,,    
\end{eqnarray}     
and the coefficients $n_{B'B}$ are given by
\begin{eqnarray}
n_{NN} = n_{D}\,+\,n_{F}\,,\,\,
n_{\Lambda\Sigma} = 2n_{D}\,,\,\,  
n_{\Xi\Xi} = n_{D}\,-\,n_{F}\,,\,\,
n_{\Sigma\Sigma} = 2n_{F}\,. 
\end{eqnarray}   
The coefficeints $C_{\phi}$, $E_{\phi}$, $\bar{E}_{\phi\phi'}$, $A_{\phi}$, 
$\bar{A}_{\phi\phi'}$, $Y_{\phi}$, $\bar{Y}_{\phi\phi'}$
are given in tables \ref{tabi1}, \ref{tabi2},
\ref{tabi3}, \ref{tabi5} and \ref{tabi7} while the 
non-analytic functions
appearing in Eq.~(\ref{matrixiso}), namely, 
${\cal L}$'s, ${\cal J}$'s, ${\cal K}$'s , ${\cal R}$'s, ${\cal T}$'s and
${\cal S}$'s are given by
\begin{eqnarray}
{\cal L}(m,\mu) = m^{2}\log\Big(\frac{m^{2}}{\mu^2}\Big) \,, \nonumber \\
\end{eqnarray}
\begin{eqnarray}
{\cal K}(m,\Delta,\mu) & = & \Big(m^2-{2\over 3}\Delta^2\Big)\log\Big({m^2\over\mu^2}\Big) \nonumber \\
&& \quad\,+\,  {2\over 3}\Delta \sqrt{\Delta^2-m^2}
\log\Big({\Delta-\sqrt{\Delta^2-m^2+ i \epsilon}\over
\Delta+\sqrt{\Delta^2-m^2+ i \epsilon}}\Big)
\nonumber\\
& & \quad \, +\, {2\over 3} {m^2\over\Delta} \Big(\ \pi m - 
\sqrt{\Delta^2-m^2}
\log\Big({\Delta-\sqrt{\Delta^2-m^2+ i \epsilon}\over
\Delta+\sqrt{\Delta^2-m^2+ i \epsilon}}\Big)
\Big)\,,
\label{eq:Kdecfun}
\end{eqnarray}
\begin{eqnarray}
{\cal J}(m,\Delta,\mu) & = & \Big(m^2-2\Delta^2\Big)\log\Big({m^2\over\mu^2}\Big) \quad \qquad \nonumber \\
&&\quad \,+\,2\Delta\sqrt{\Delta^2-m^2}
\log\Big({\Delta-\sqrt{\Delta^2-m^2+ i \epsilon}\over
\Delta+\sqrt{\Delta^2-m^2 + i \epsilon}}\Big)\,,
\label{eq:decfun}
\end{eqnarray}
\begin{eqnarray} 
{\cal R}(\eta_{\phi}\eta_{\phi'},\Delta,\mu) &=& {\cal H}({\cal
  L}(m_{\eta_{\phi}},\mu),{\cal L}(m_{\eta_{\phi'}},\mu),
{\cal L}(m_{X},\mu))\,,\nonumber \\ 
{\cal T}(\eta_{\phi}\eta_{\phi'},\Delta,\mu) &=& {\cal H}({\cal J}(m_{\eta_{\phi}},\Delta,\mu),
{\cal J}(m_{\eta_{\phi'}},\Delta,\mu),{\cal J}(m_{X},\Delta,\mu)\,,\, \nonumber \\
{\cal S}(\eta_{\phi}\eta_{\phi'},\Delta,\mu) &=& {\cal H}({\cal K}(m_{\eta_{\phi}},\Delta,\mu),{\cal K}(m_{\eta_{\phi'}}\Delta,\mu),
{\cal K}(m_{X},\Delta,\mu))\,.
\label{hairpinoneloop} 
\end{eqnarray}
At NLO, the momentum behavior of the axial form factors
is purely polynomial. This in turn implies that the axial
radii $\langle r^{2}_{B'B} \rangle$ which are defined by $ \langle r^{2}_{B'B} 
\rangle \equiv \lim_{q \rightarrow 0}6\frac{d}{dq^2}G_{A,B'B}(q^2)$, 
are insensitive to NLO chiral corrections. Since
no $q^2$ dependence appears in the non-analytic functions from one-loop
diagrams, the axial form factors are insensitive to the long-range 
effects introduced by boundary conditions. 
Therefore, flavor twisted boundary conditions can be used to produce momentum
transfer between initial and final baryon states without sizable finite 
volume corrections to the extraction of the axial radii \cite{twisted13}.      
 
The one-loop diagrams which contribute at NLO to the pseudoscalar
form factor are depicted in Fig.2. Additionally there are further
diagrams generated by the insertion of local interactions from
the fourth-order meson Lagrangian. Despite the large number of
diagrams, there are a number of simplifications. In particular
the second and fourth diagrams of the second line in Fig. 2
(along with local insertions on the meson line) lead to the one-loop
renormalized pion propagator. Additionally the second and fourth
diagrams of the first row in Fig 2 (along with NLO pion axial coupling)
contribute to the one-loop value of the pion decay constant.
The remaining ten diagrams are generated from vertices in the NLO
baryon Lagrangian. These diagrams renormalize the tree-level axial
coupling of the pion to the baryons. Carefully accounting for all of 
these factors, we find
\begin{eqnarray}
G_{P,B'B}(q^2)\,& = &\,(m_B+m_{B'})^2
 \Big(\,\frac{f_{\pi}/f}{q^2-m^2_{\pi}}G_{A,B'B}(0)\sqrt{Z_{\pi}}\,\,
-\, \frac{1}{3}\langle r^{2}_{B'B} \rangle\,\Big)\,,
\label{pseudoiso}
\end{eqnarray}
where in this NLO expression, the axial charge $G_{A,B'B}(0)$,
pion mass $m_{\pi}$ and pion decay constant $f_\pi$ are taken to be 
their physical values and $Z_{\pi}$ is the pion wavefunction
renormalization which is shown in the Appendix. In fitting lattice
data to Eq.~(\ref{pseudoiso}), one would thus use the lattice measured 
values for $G_{A,B'B}(0)$, $m_{\pi}$ and $f_{\pi}$.  The final 
term in the pseudoscalar form factor is $\langle r^{2}_{B'B} \rangle$, 
which is the axial radius. Its
appearance here was discovered long ago under the guise of PCAC by
Adler and Dothan \cite{Adler66}

The simple structure of the psuedoscalar form factor at NLO
in both \CPT\ and \PQCPT\, allows one perform an approximate check
of the Goldberger-Treiman relation. The residue of the pseudoscalar
form factor at the pion pole is proportional to the pion-baryon-baryon
coupling $G_{\pi B'B}$.
One can thus perturbatively investigate the Goldberger-Treiman
relation using a lattice determination of the pseudoscalar form
factor. This indirect method is considerably simpler
than a lattice measurement of baryon-to-baryon-plus-pion correlation
functions which contain final state interactions.

\begin{table}[!ht]
\begin{center}
\begin{tabular}{cc}
\multicolumn{2}{c}{$C_{\phi}$}\tabularnewline
\hline
$uj$&
$ur$\tabularnewline
$-2$&
$-1$\tabularnewline
\end{tabular}
\end{center}
\caption{The coefficients $C_{\phi}$ 
in PQ$\chi$PT for the isospin changing axial form factors. 
The $C_{\phi}$ are categorized by the loop mesons $\phi$ with 
mass $m_{\phi}$ and are the same for all isospin transitions.}
\label{tabi1}
\end{table}
\begin{table}[!ht]
\begin{center}
\begin{tabular}{cccccccc|ccc}
\multicolumn{8}{c|}{$E_{\phi}$}&
\multicolumn{3}{c}{$\bar{E}_{\phi\phi'}$}\tabularnewline
\cline{1-8} 
\hline
&
$\eta_{u}$&
$\eta_{s}$&
$us$&
$uj$&
$ur$&
$sj$&
$sr$&
$\eta_{u}\eta_{u}$&
$\eta_{u}\eta_{s}$&
$\eta_{s}\eta_{s}$\tabularnewline
$NN$&
$-\frac{20}{27}$&
$0$&
$0$&
$-\frac{40}{81}$&
$-\frac{20}{81}$&
$0$&
$0$&
$0$&
$0$&
$0$\tabularnewline
$\Lambda\Sigma$&
$-\frac{10}{27}$&
$0$&
$-\frac{10}{54}$&
$-\frac{10}{27}$&
$-\frac{10}{54}$&
$0$&
$0$&
$0$&
$0$&
$0$\tabularnewline
$\Xi\Xi$&
$0$&
$\frac{5}{81}$&
$\frac{10}{81}$&
$0$&
$0$&
$\frac{10}{81}$&
$\frac{5}{81}$&
$\frac{10}{81}$&
$-\frac{20}{81}$&
$\frac{10}{81}$\tabularnewline
$\Sigma\Sigma$&
$-\frac{10}{81}$&
$0$&
$-\frac{65}{81}$&
$-\frac{10}{81}$&
$-\frac{5}{81}$&
$-\frac{40}{81}$&
$-\frac{20}{81}$&
$-\frac{20}{81}$&
$\frac{40}{81}$&
$-\frac{20}{81}$\tabularnewline
\end{tabular}
\end{center}
\caption{The coefficients $E_{\phi}$ and $\bar{E}_{\phi\phi'}$ 
in PQ$\chi$PT for the isospin changing axial form factors. 
The $E_{\phi}$ are categorized by the loop mesons $\phi$ with 
mass $m_{\phi}$ and $\bar{E}_{\phi\phi'}$ are listed
by pairs $\phi \phi'$ of $\eta_q$ mesons.}
\label{tabi2}
\end{table}

\begin{table}[!ht]
\begin{center}
\begin{tabular}{ccccc|cc}
\multicolumn{5}{c|}{$A_{\phi}$}&
{$\bar{A}_{\phi\phi'}$}\tabularnewline 
\hline 
&
$\eta_{u}$ & $\eta_{s}$ & $us$ & $uj$ & $\eta_{u}\eta_{u}$ & $\eta_{u}\eta_{s}$\tabularnewline
$NN$ & $\frac{8}{9}D+\frac{8}{3}F$ & $0$ & $0$ & $\frac{8}{3}D+\frac{8}{9}F$
& $0$ & $0$\tabularnewline
$\Lambda\Sigma$ & $-\frac{4}{9}D+\frac{4}{3}F$ & $0$ & $\frac{16}{9}D+\frac{8}{3}F$ & 
 $\frac{16}{9}D$ & $\frac{16}{9}D-\frac{8}{3}F$ & $-\frac{20}{9}D+\frac{4}{3}F$\tabularnewline
$\Xi\Xi$ & $0$ & $\frac{8}{9}F$ & $\frac{4}{9}D+\frac{4}{9}F$ 
& $0$ & $\frac{8}{9}D-\frac{8}{9}F$ & $-\frac{8}{9}D-\frac{8}{9}F$\tabularnewline
$\Sigma\Sigma$ & $\frac{8}{9}F$ & $0$ & $\frac{4}{9}D+\frac{4}{9}F$ &
$\frac{8}{9}D+\frac{8}{9}F$ & $\frac{16}{9}F$ & $-\frac{8}{9}D-\frac{8}{9}F$ \tabularnewline
\hline 
\hline
&
$ur$ & $sj$ & $sr$ & $$ & $\eta_{s}\eta_{s}$ & $$\tabularnewline
$NN$ & $\frac{4}{3}D+\frac{4}{9}F$ & $0$ & $0$ & $$
& $0$ & $$\tabularnewline
$\Lambda\Sigma$ & $\frac{8}{9}D$ & $\frac{8}{9}D+\frac{8}{3}F$ & 
$\frac{4}{9}D+\frac{4}{3}F$ & $$ & $\frac{4}{9}D+\frac{4}{3}F$ & $$\tabularnewline
$\Xi\Xi$ & $0$ & $\frac{16}{9}F$ & $\frac{8}{9}F$
& $$ & $\frac{16}{9}F$ & $$\tabularnewline
$\Sigma\Sigma$ & $\frac{4}{9}D+\frac{4}{9}F$ &
$\frac{16}{9}D-\frac{16}{9}F$ & 
$\frac{8}{9}D-\frac{8}{9}F$ & $$ & $\frac{8}{9}D-\frac{8}{9}F$ & $$ \tabularnewline
\end{tabular}
\end{center}
\caption{The coefficients $A_{\phi}$ and $\bar{A}_{\phi\phi'}$ 
in PQ$\chi$PT for the isospin changing axial form factors. 
The $A_{\phi}$ and $\bar{A}_{\phi\phi'}$ coefficients are categorized
as in Table \ref{tabi2}.}
\label{tabi3}
\end{table}

\begin{table}[!ht]
\begin{center}
\begin{tabular}{ccc|c}
\multicolumn{3}{c|}{$Y_{\phi}$}&
{$\bar{Y}_{\phi\phi'}$}\tabularnewline 
\hline 
&
$\eta_{u}$ & $\eta_{s}$ & $\eta_{u}\eta_{u}$\tabularnewline
$NN$ & $-\frac{4}{3}D^{3}+\frac{16}{3}D^{2}F-4DF^{2}$ & $0$ & $-D^{3}+5D^{2}F-3DF^{2}-9F^{3}$\tabularnewline
$\Lambda\Sigma$ & $-\frac{16}{9}D^{3}+\frac{16}{3}D^{2}F$ & $0$ & $\frac{16}{3}D^{2}F-8DF^{2}$\tabularnewline
$\Xi\Xi$ & $0$ & $-\frac{2}{9}D^{3}-\frac{2}{3}D^{2}F-2DF^{2}+2F^{3}$ & $-D^{3}+3D^{2}F-3DF^{2}+F^{3}$\tabularnewline
$\Sigma\Sigma$ & $-\frac{8}{9}D^{3}+\frac{4}{3}D^{2}F-4F^{3}$ & $0$ & $-8F^{3}$\tabularnewline
\hline 
\hline
&
$us$ & $uj$ & $\eta_{u}\eta_{s}$\tabularnewline
$NN$ & $0$ & $\frac{4}{3}D^3-\frac{4}{3}D^2F+4DF^2-4F^3$ & $0$\tabularnewline
$\Lambda\Sigma$ & $\frac{4}{9}D^3-\frac{4}{3}D^2F$ & 
$\frac{4}{9}D^3-\frac{8}{3}D^2F+4DF^2$ & $-\frac{8}{3}D^3+\frac{16}{3}D^2F-8DF^2$\tabularnewline
$\Xi\Xi$ & $\frac{2}{9}D^3-\frac{2}{3}D^2F+6DF^2-2F^3$ & 
$0$ & $4D^2F-8DF^2+4F^3$\tabularnewline
$\Sigma\Sigma$ & $-\frac{4}{9}D^3+\frac{16}{3}D^2F-8DF^2+4F^3$ & 
$\frac{8}{9}D^3+\frac{4}{3}D^2F-4F^3$ & $8DF^2-8F^3$\tabularnewline
\hline 
\hline
&
$ur$ & $sj$ & $\eta_{s}\eta_{s}$\tabularnewline
$NN$ & $\frac{2}{3}D^3-\frac{2}{3}D^2F+2DF^2-2F^3$ & 
$0$ & $0$\tabularnewline
$\Lambda\Sigma$ & $\frac{2}{9}D^3-\frac{4}{3}D^2F+2DF^2$ & 
$\frac{4}{3}D^3+\frac{8}{3}D^2F-4DF^2$ & $\frac{2}{3}D^3+\frac{4}{3}D^2F-2DF^2$\tabularnewline
$\Xi\Xi$ & $0$ & 
$\frac{4}{9}D^3+\frac{4}{3}D^2F-4DF^2+4F^3$ & $-4DF^2+4F^3$\tabularnewline
$\Sigma\Sigma$ & $\frac{4}{9}D^3+\frac{2}{3}D^2F-2F^3$ & 
$-4D^2F+8DF^2-4F^3$ & $-2D^2F+4DF^2-2F^3$\tabularnewline
\hline 
\hline
&
$sr$ & $$ & $$\tabularnewline
$NN$ & $0$ & 
$$ & $$\tabularnewline
$\Lambda\Sigma$ & $\frac{2}{3}D^3+\frac{4}{3}D^2F-2DF^2$ & 
$$ & $$\tabularnewline
$\Xi\Xi$ & $\frac{2}{9}D^3+\frac{2}{3}D^2F-2DF^2+2F^3$ & 
$$ & $$\tabularnewline
$\Sigma\Sigma$ & $-2D^2F+4DF^2-2F^3$ & 
$$ & $$\tabularnewline
\end{tabular}
\end{center}
\caption{The coefficients $Y_{\phi}$ and $\bar{Y}_{\phi\phi'}$ 
in PQ$\chi$PT for the isospin changing axial form factors. 
The $Y_{\phi}$ and $\bar{Y}_{\phi\phi'}$ coefficients are categorized 
as in Table \ref{tabi2}.}
\label{tabi5}
\bigskip
\bigskip
\end{table}

\begin{table}[!ht]
\begin{center}
\begin{tabular}{ccccc}
\multicolumn{5}{c}{$u_{\phi}$}\tabularnewline
\hline
&
$\,\,\,uu\,\,\,\,$&
$\,\,\,ss\,\,\,\,$&
$\,\,\,jj\,\,\,\,$&
$\,\,\,rr\,\,\,\,$\tabularnewline
$\,\,\,NN\,\,\,\,$&
$\,\,\,\,-\frac{1}{3}b_{1}+\frac{2}{3}b_2-\frac{1}{6}b_3+\frac{1}{6}b_4
+\frac{1}{3}b_5\,\,\,\,$&
$\,\,\,\,0\,\,\,\,$&
$\,\,\,\,\frac{2}{3}b_7-\frac{1}{3}b_6\,\,\,\,$&
$\,\,\,\,\frac{1}{3}b_7-\frac{1}{6}b_6\,\,\,$\tabularnewline
$\,\,\,\Lambda\Sigma\,\,\,\,$&
$\,\,\,\,-b_{1}+\frac{1}{2}b_2-\frac{1}{4}b_3
+\frac{1}{4}b_5\,\,\,\,$&
$\,\,\,\,-\frac{1}{4}b_3+\frac{1}{2}b_4\,\,\,\,$&
$\,\,\,\,\frac{1}{2}b_7-b_6\,\,\,\,$&
$\,\,\,\,\frac{1}{4}b_7-\frac{1}{2}b_6\,\,\,$\tabularnewline
$\,\,\,\Xi\Xi\,\,\,\,$&
$\,\,\,\,-\frac{2}{3}b_{1}-\frac{1}{6}b_2\,\,\,\,$&
$\,\,\,\,-\frac{1}{3}b_3+\frac{1}{3}b_4-\frac{1}{12}b_5\,\,\,\,$&
$\,\,\,\,-\frac{1}{6}b_7-\frac{2}{3}b_6\,\,\,\,$&
$\,\,\,\,-\frac{1}{12}b_7-\frac{1}{3}b_6\,\,\,$\tabularnewline
$\,\,\,\Sigma\Sigma\,\,\,\,$&
$\,\,\,\,\frac{1}{3}b_{1}+\frac{5}{6}b_2+\frac{1}{12}b_3+\frac{1}{6}b_4
+\frac{1}{12}b_5\,\,\,\,$&
$\,\,\,\,\frac{1}{12}b_3-\frac{1}{3}b_4+\frac{1}{3}b_5\,\,\,\,$&
$\,\,\,\,\frac{5}{6}b_7+\frac{1}{3}b_6\,\,\,\,$&
$\,\,\,\,\frac{5}{12}b_7+\frac{1}{6}b_6\,\,\,$\tabularnewline
\end{tabular}
\end{center}
\caption{The coefficients $u_{\phi}$ 
in PQ$\chi$PT for the isospin changing axial form factors. The 
$u_{\phi}$  coefficients are categorized by the mesons with 
mass $m_{\phi}$.}
\label{tabi7}
\bigskip
\bigskip
\end{table}

\subsection{Strangeness Changing Transitions}
\label{SCT}
The $\Delta S = -1$ strangeness changing transitions corresponds to
the flavor matrix $\overline{\lambda}^{4+5i}$ which is given by
\begin{equation} \label{eqn:tau1}
\overline{\lambda}^{4+5i}_{i j} = 
\begin{cases}
 1 \quad i = 1\, , j = 3 \\
 0  \quad {\text{otherwise}}
\end{cases}
\,,\end{equation}
in the partially quenched theory. With Eq.~(\ref{eqn:tau1}) there exists six 
strangeness changing transitions among the hyperons
\begin{equation}
\Sigma^{0} \rightarrow p\,\,,\,\,\Sigma^{-} \rightarrow n\,\,,\,\,
\Lambda \rightarrow p\,\,,\,\,\Xi^{0}
\rightarrow \Sigma^{+}\,\,,\,\,
\Xi^{-} \rightarrow \Sigma^{0}\,\,,\,\,\Xi^{-} \rightarrow \Lambda\,\,.
\end{equation}
The $N$$\Lambda$ transition matrix elements define the $N$$\Lambda$ 
axial form factor $G_{A,N\Lambda}(q^2)$ and induced pseudoscalar form factor
$G_{P,N\Lambda}(q^2)$
\begin{equation}
\langle p (P')| J^{4+5i}_{\mu,5} | \Lambda (P) \rangle\,=\,-\frac{1}{\sqrt{6}}
\overline{U}_{N}(P')\Big[ 2S_{\mu}G_{A,N\Lambda}(q^2)+\frac{q_{\mu} q \cdot
  S}{(m_N+m_{\Lambda})^2}G_{P,N\Lambda}(q^2)\Big]U_{\Lambda}(P)\,,
\end{equation}
where, as above $q_{\mu} = (P' - P)_{\mu}$ is the four momentum transfer. The
$\Lambda$$\Xi$ transition matrix elements define the 
$\Lambda$$\Xi$ axial form factor $G_{A,\Lambda\Xi}$ and induced pseudoscalar
form factor $G_{P,\Lambda\Xi}$
\begin{equation}
\langle \Lambda (P') | J^{4+5i}_{\mu,5}| \Xi^{-} (P)\rangle\,=\,
\frac{1}{\sqrt{6}} 
\overline{U}_{\Lambda}(P')\Big[2S_{\mu}G_{A,\Lambda\Xi}(q^2)
+ \frac{q_{\mu} q \cdot S}{(m_{\Lambda}+m_{\Xi})^2}G_{P,\Lambda\Xi}(q^2)\Big]U_{\Xi}(P)\,.
\end{equation}
The $N$$\Sigma$ axial transitions defines the $N$$\Sigma$ axial form factor 
$G_{A,N\Sigma}(q^2)$ and the induced pseudoscalar form factor $G_{P,N\Sigma}(q^2)$
\begin{equation}
\langle n (P') | J^{+}_{\mu,5} | \Sigma^{-} (P)\rangle\,=\, 
\overline{U}_{N}(P')\Big[2S_{\mu}G_{A,N\Sigma}(q^2) + \frac{q_{\mu} q \cdot
  S}{(m_N+m_{\Sigma})^2}G_{P,N\Sigma}(q^2)\Big]U_{\Sigma}(P)\,.
\end{equation}
Finally, the $\Sigma$$\Xi$ axial form factor $G_{A,\Sigma\Xi}(q^2)$ and
the induced pseudoscalar form factor $G_{P,\Sigma\Xi}(q^2)$ is defined 
through the $\Sigma$$\Xi$  transition matrix element
\begin{equation}
\langle \Sigma^{0} (P') | J^{4+5}_{\mu,5}| \Xi^{-} (P)\rangle\,=\,
\frac{1}{\sqrt{2}}\overline{U}_{\Sigma}(P')\Big[2S_{\mu}G_{A,\Sigma\Xi}(q^2)+
\frac{q_{\mu} q \cdot S}{(m_{\Sigma}+m_{\Xi})^2}G_{P,\Sigma\xi}(q^2)\Big]U_{\Xi}(P)\,.
\end{equation} 
Notice while there are two $N$$\Sigma$ and two $\Sigma$$\Xi$ transitions, 
both of their matrix elements are 
related by isospin factors, namely:
\begin{eqnarray}
\langle p (P')| J^{4+5i}_{\mu,5} | 
\Sigma^{0}(P)\rangle\,&=&\,
\frac{1}{\sqrt{2}}\langle n (P')| J^{4+5i}_{\mu,5} | \Sigma^{-}(P)\rangle
\,, \nonumber \\
\langle \Sigma^{0} (P') | J^{4+5i}_{\mu,5} | \Xi^{-} (P)\rangle\,&=&\,\frac{1}{\sqrt{2}}
\langle \Sigma^{+} (P')| J^{4+5i}_{\mu,5} | \Xi^{0} (P)\rangle\,.
\end{eqnarray}
Following the same considerations in section \ref{IST} and 
assembling the LO and NLO  contributions, the axial form factors of strangeness
changing transitions are given by
\begin{eqnarray}
G_{A,B'B}(q^2) & = &\,  g_{B'B}\sqrt{Z_{B'}Z_{B}} + \frac{1}{16\pi^2f^2}\Bigg[ g_{B'B}\Big(
\sum_{\phi}C_{\phi}{\cal L}(m_{\phi},\mu)
\nonumber \\
& + &\sum_{\phi\phi^{\prime}} \bar{C}_{\phi\phi^{\prime}}{\cal
  R}(\eta_{\phi}\eta_{\phi'},\Delta,\mu)\Big)
+ {\cal H}{\cal C}^2\Big(\sum_{\phi} E_{\phi}{\cal J}(m_{\phi},\Delta,\mu) \nonumber \\
& + &\sum_{\phi\phi^{\prime}} \bar{E}_{\phi\phi^{\prime}}{\cal T}(\eta_{\phi}
  \eta_{\phi'},\Delta,\mu)\Big) \nonumber \\
& + & \cC^2 \Big( \sum_{\phi}A_{\phi}{\cal K}(m_{\phi},\Delta,\mu)
 + \sum_{\phi\phi^{\prime}}
\bar{A}_{\phi\phi^{\prime}}
{\cal S}(\eta_{\phi} \eta_{\phi^{\prime}},\Delta,\mu) \Big) \nonumber \\
& + &\sum_{\phi} Y_{\phi}{\cal L}(m_{\phi},\mu)
 + \sum_{\phi\phi^{\prime}} \bar{Y}_{\phi,\phi^{\prime}}{\cal
  R}(\eta_{\phi}\eta_{\phi'},\Delta,\mu) \Bigg] \nonumber \\
& + & n_{B'B}\frac{q^2}{\Lambda_{\chi}^2} + \sum_{\phi}u_{\phi}m_{\phi}^2
\,.  
\label{matrixstrange}
\end{eqnarray}
In Eq.~(\ref{matrixstrange}), we use $B$ and
$B'$ to denote the initial and final states of octet baryon, $Z_B$ and
$Z_{B'}$ are again the wave function renormalization for which the explicit 
expressions are given in the Appendix. The $g_{B'B}$'s appearing
above are the leading order octet baryon axial charges
\begin{eqnarray}
g_{N\Lambda} &=& (3F\,+\,D)\,,\,\,\, 
g_{\Lambda\Xi} = (3F\,-\,D)\,, 
\nonumber \\
g_{N\Sigma} &=& (D\,-\,F)\,,\,\,\,g_{\Sigma\Xi} = (F\,+\,D),
\end{eqnarray} 
and the coefficients $n_{B'B}$ are given by a formula of exactly the same form
\begin{eqnarray}
n_{N\Lambda} &=& 3n_{F}\,+\,n_{D}\,,\,\,
n_{\Lambda\Xi} = 3n_{F}\,-\,n_{D}\,,\nonumber \\
n_{N\Sigma} &=& n_{D}\,-\,n_{F}\,,\,\,
n_{\Sigma\Xi} = n_{D}\,+\,n_{F}\,.    
\end{eqnarray}   
The coefficeints $C_{\phi}$, $\bar{C}_{\phi\phi'}$, $E_{\phi}$, $\bar{E}_{\phi\phi'}$, $A_{\phi}$, 
$\bar{A}_{\phi\phi'}$, $Y_{\phi}$, $\bar{Y}_{\phi\phi'}$
are given in tables \ref{tabs1}, \ref{tabs2},
\ref{tabs3}, \ref{tabs5}. 
\begin{table}[!ht]
\begin{center}
\begin{tabular}{cccc|ccc}
\multicolumn{4}{c|}{$C_{\phi}$}&
\multicolumn{3}{c}{$\bar{C}_{\phi\phi'}$}\tabularnewline
\hline
$uj$&
$ur$&
$sj$&
$sr$&
$\eta_{u}\eta_{u}$&
$\eta_{u}\eta_{s}$&
$\eta_{s}\eta_{s}$\tabularnewline
$-1$&
$-\frac{1}{2}$&
$-1$&
$-\frac{1}{2}$&
$-\frac{1}{2}$&
$1$&
$-\frac{1}{2}$\tabularnewline
\end{tabular}
\end{center}
\caption{The coefficients $C_{\phi}$ and $\bar{C}_{\phi\phi'}$ 
in PQ$\chi$PT, which are all the same for the strangeness changing 
axial form factors. The $C_{\phi}$ and $\bar{C}_{\phi\phi'}$ coefficients
are categorized as in Table \ref{tabi2}.}
\label{tabs1}
\end{table}
\begin{table}[!ht]
\begin{center}
\begin{tabular}{cccccccc|ccc}
\multicolumn{8}{c|}{$E_{\phi}$}&
\multicolumn{3}{c}{$\bar{E}_{\phi\phi'}$}\tabularnewline
\hline
&
$\eta_{u}$&
$\eta_{s}$&
$us$&
$uj$&
$ur$&
$sj$&
$sr$&
$\eta_{u}\eta_{u}$&
$\eta_{u}\eta_{s}$&
$\eta_{s}\eta_{s}$\tabularnewline
$N\Lambda$&
$-\frac{10}{9}$&
$0$&
$-\frac{10}{18}$&
$-\frac{10}{9}$&
$-\frac{10}{18}$&
$0$&
$0$&
$0$&
$0$&
$0$\tabularnewline
$\Lambda\Xi$&
$-\frac{10}{27}$&
$0$&
$-\frac{20}{27}$&
$-\frac{20}{27}$&
$-\frac{10}{27}$&
$0$&
$0$&
$0$&
$0$&
$0$\tabularnewline
$N\Sigma$&
$\frac{10}{81}$&
$0$&
$\frac{5}{81}$&
$\frac{10}{81}$&
$\frac{5}{81}$&
$0$&
$0$&
$0$&
$0$&
$0$\tabularnewline
$\Sigma\Xi$&
$-\frac{10}{81}$&
$-\frac{10}{81}$&
$-\frac{40}{81}$&
$-\frac{20}{81}$&
$-\frac{10}{81}$&
$-\frac{20}{81}$&
$-\frac{10}{81}$&
$-\frac{20}{81}$&
$\frac{40}{81}$&
$-\frac{20}{81}$\tabularnewline
\end{tabular}
\end{center}
\caption{The coefficients $E_{\phi}$ and $\bar{E}_{\phi\phi'}$ 
in PQ$\chi$PT for the strangeness changing axial form factors. 
The $E_{\phi}$ and $\bar{E}_{\phi\phi'}$ coefficients are categorized as in
Table \ref{tabi2}.}
\label{tabs2}
\end{table}
%
%
\begin{table}[!ht]
\begin{center}
\begin{tabular}{ccccc|cc}
\multicolumn{5}{c|}{$A_{\phi}$}&
{$\bar{A}_{\phi\phi'}$}\tabularnewline 
\hline 
&
$\eta_{u}$ & $\eta_{s}$ & $us$ & $uj$ & $\eta_{u}\eta_{u}$ & $\eta_{u}\eta_{s}$\tabularnewline
$N\Lambda$ & $2D+2F$ & $0$ & $-\frac{2}{3}D+2F$ & $\frac{16}{3}D$
& $0$ & $0$\tabularnewline
$\Lambda\Xi$ & $-\frac{2}{3}D+\frac{2}{3}F$ & $\frac{2}{9}D+\frac{2}{3}F$
& $\frac{4}{9}D-\frac{4}{3}F$ & 
$\frac{20}{9}D-4F$ & $\frac{16}{9}D-\frac{8}{3}F$ & $-\frac{20}{9}D+\frac{4}{3}F$\tabularnewline
$N\Sigma$ & $\frac{2}{9}D+\frac{10}{9}F$ & $0$ & $\frac{2}{9}D+\frac{2}{9}F$
& $\frac{16}{9}F$ & $-\frac{4}{9}D+\frac{4}{3}F$ & $\frac{4}{9}D-\frac{4}{3}F$\tabularnewline
$\Sigma\Xi$ & $-\frac{2}{9}D+\frac{2}{9}F$ &
$-\frac{2}{9}D+\frac{2}{9}F$ & $\frac{4}{3}D+\frac{20}{9}F$ &
$\frac{4}{3}D+\frac{4}{9}F$ & $\frac{4}{9}D+\frac{4}{9}F$ & $-\frac{8}{9}D-\frac{8}{9}F$ \tabularnewline
\hline 
\hline
&
$ur$ & $sj$ & $sr$ & $$ & $\eta_{s}\eta_{s}$ & $$\tabularnewline
$N\Lambda$ & $\frac{8}{3}D$ & $0$ & $0$ & $$
& $0$ & $$\tabularnewline
$\Lambda\Xi$ & $\frac{10}{9}D-2F$ & $\frac{4}{9}D+\frac{4}{3}F$ & 
$\frac{2}{9}D+\frac{2}{3}F$ & $$ & $\frac{4}{9}D+\frac{4}{3}F$ & $$\tabularnewline
$N\Sigma$ & $\frac{8}{9}F$ & $0$ & $0$
& $$ & $0$ & $$\tabularnewline
$\Sigma\Xi$ & $\frac{2}{3}D+\frac{2}{9}F$ &
$\frac{4}{3}D+\frac{4}{9}F$ & 
$\frac{2}{3}D+\frac{2}{9}F$ & $$ & $\frac{4}{9}D+\frac{4}{9}F$ & $$ \tabularnewline
\end{tabular}
\end{center}
\caption{The coefficients $A_{\phi}$ and $\bar{A}_{\phi\phi'}$ 
in PQ$\chi$PT for the strangeness changing axial form factors. 
The $A_{\phi}$ and $\bar{A}_{\phi\phi'}$ coefficients are categorized as
in Table \ref{tabi2}.}
\bigskip
\label{tabs3}
\end{table}
\begin{table}[!ht]
\begin{center}
\begin{tabular}{ccc|c}
\multicolumn{3}{c|}{$Y_{\phi}$}&
{$\bar{Y}_{\phi\phi'}$}\tabularnewline 
\hline 
&
$\eta_{u}$ & $\eta_{s}$ & $\eta_{u}\eta_{u}$\tabularnewline
$N\Lambda$ & $\frac{1}{9}D^3+5D^2F-9DF^2+3F^3$ & $0$ & $-\frac{4}{3}D^3+2D^2F+12DF^2-18F^3$\tabularnewline
$\Lambda\Xi$ &
$-\frac{5}{3}D^3+\frac{11}{3}D^2F-5DF^2+3F^3$ & $-\frac{5}{9}D^3-D^2F+DF^2-3F^3$ &
$\frac{4}{3}D^3-\frac{22}{3}D^2F+12DF^2-6F^3$\tabularnewline
$N\Sigma$ & $-\frac{1}{9}D^3-D^2F+DF^2+F^3$ & 
$0$ & $2D^2F-8DF^2+6F^3$\tabularnewline
$\Sigma\Xi$ & $-\frac{1}{3}D^3+\frac{7}{3}D^2F-DF^2-F^3$ & 
$-\frac{1}{3}D^3+\frac{7}{3}D^2F-DF^2-F^3$ & $2D^2F-2F^{3}$\tabularnewline
\hline 
\hline
&
$us$ & $uj$ & $\eta_{u}\eta_{s}$\tabularnewline
$N\Lambda$ & $-\frac{25}{9}D^3+7D^2F-3DF^2-3F^3$ &
$\frac{20}{9}D^3-4D^2F+12DF^2-12F^3$ & 
$\frac{1}{3}D^3+D^2F-3DF^2-9F^3$\tabularnewline
$\Lambda\Xi$ & $\frac{8}{9}D^3+\frac{16}{3}D^2F-8DF^2$ & 
$-\frac{2}{3}D^3-\frac{22}{3}D^2F+14DF^2-6F^3$ & $-\frac{1}{3}D^3-\frac{7}{3}D^2F+15DF^2-15F^3$\tabularnewline
$N\Sigma$ & $\frac{1}{9}D^3-\frac{1}{3}D^2F+3DF^2-F^3$ & 
$\frac{4}{9}D^3+\frac{4}{3}D^2F-4DF^2+4F^3$ & $-D^3+5D^2F-7DF^2+3F^3$\tabularnewline
$\Sigma\Xi$ & $-\frac{2}{3}D^3+\frac{2}{3}D^2F-2DF^2+2F^3$ & 
$\frac{2}{3}D^3-\frac{2}{3}D^2F+2DF^2-2F^3$ & $-D^3+D^2F-3DF^2-5F^3$\tabularnewline
\hline 
\hline
&
$ur$ & $sj$ & $\eta_{s}\eta_{s}$\tabularnewline
$N\Lambda$ & $\frac{10}{9}D^3-2D^2F+6DF^2-6F^3$ & 
$0$ & $0$\tabularnewline
$\Lambda\Xi$ & $-\frac{1}{3}D^3-\frac{11}{3}D^2F+7DF^2-3F^3$ & 
$\frac{10}{9}D^3+\frac{10}{3}D^2F-2DF^2-6F^3$ & $\frac{2}{3}D^2F-6F^3$\tabularnewline
$N\Sigma$ & $\frac{2}{9}D^3+\frac{2}{3}D^2F-2DF^2+2F^3$ & 
$0$ & $0$\tabularnewline
$\Sigma\Xi$ & $\frac{1}{3}D^3-\frac{1}{3}D^2F+DF^2-F^3$ & 
$\frac{2}{3}D^3-\frac{2}{3}D^2F+2DF^2-2F^3$ & $2D^2F-2F^3$\tabularnewline
\hline 
\hline
&
$sr$ & $$ & $$\tabularnewline
$N\Lambda$ & $0$ & 
$$ & $$\tabularnewline
$\Lambda\Xi$ & $\frac{5}{9}D^3+\frac{5}{3}D^2F-DF^2-3F^3$ & 
$$ & $$\tabularnewline
$N\Sigma$ & $0$ & 
$$ & $$\tabularnewline
$\Sigma\Xi$ & $\frac{1}{3}D^3-\frac{1}{3}D^2F+DF^2-F^3$ & 
$$ & $$\tabularnewline
\end{tabular}
\end{center}
\caption{The coefficients $Y_{\phi}$ and $\bar{Y}_{\phi\phi'}$ 
in PQ$\chi$PT for the strangeness changing axial form factors. 
The $Y_{\phi}$ and $\bar{Y}_{\phi\phi'}$ coefficients are categorized as
in Table \ref{tabi2}.}
\label{tabs5}
\end{table}
\begin{table}[!ht]
\begin{center}
\begin{tabular}{ccccc}
\multicolumn{5}{c}{$u_{\phi}$}\tabularnewline
\hline
&
$uu\,\,$&
$ss\,\,$&
$jj\,\,$&
$rr\,\,$\tabularnewline
$N\Lambda\,\,$&
$\,\,\,\,\frac{3}{4}b_{2}+\frac{3}{4}b_5\,\,\,\,$&
$\,\,\,\,\frac{3}{4}b_2\,\,\,\,$&
$\,\,\,\,\frac{3}{2}b_7\,\,\,\,$&
$\,\,\,\,\frac{3}{4}b_7$\tabularnewline
$\Lambda\Xi\,\,$&
$\,\,\,\,\frac{1}{2}b_{1}+\frac{1}{2}b_2+\frac{1}{4}b_3-\frac{3}{4}b_4
+\frac{1}{2}b_5\,\,\,\,$&
$\,\,\,\,\frac{1}{2}b_1+\frac{1}{2}b_2+\frac{1}{4}b_3+\frac{1}{4}b_4\,\,\,\,$&
$\,\,\,\,b_7+b_6\,\,\,\,$&
$\,\,\,\,\frac{1}{2}b_7+\frac{1}{2}b_6$\tabularnewline
$N\Sigma\,\,$&
$\,\,\,\,-\frac{1}{3}b_{1}-\frac{1}{12}b_2-\frac{1}{3}b_3
+\frac{1}{3}b_4-\frac{1}{12}b_5\,\,\,\,$&
$\,\,\,\,-\frac{1}{3}b_1-\frac{1}{12}b_2\,\,\,\,$&
$\,\,\,\,-\frac{1}{6}b_7-\frac{2}{3}b_6\,\,\,\,$&
$\,\,\,\,-\frac{1}{12}b_7-\frac{1}{3}b_6$\tabularnewline
$\Sigma\Xi\,\,$&
$\,\,\,\,-\frac{1}{6}b_{1}+\frac{1}{3}b_2-\frac{1}{12}b_3+\frac{1}{12}b_4
+\frac{1}{6}b_5\,\,\,\,$&
$\,\,\,\,-\frac{1}{6}b_{1}+\frac{1}{3}b_2-\frac{1}{12}b_3+\frac{1}{12}b_4
+\frac{1}{6}b_5\,\,\,\,$&
$\,\,\,\,\frac{2}{3}b_7-\frac{1}{3}b_6\,\,\,\,$&
$\,\,\,\,\frac{1}{3}b_7-\frac{1}{6}b_6$\tabularnewline
\end{tabular}
\end{center}
\caption{The coefficients $u_{\phi}$ 
in PQ$\chi$PT for the strangeness changing axial 
form factors. The coefficients $u_{\phi}$ are 
categorized by the mesons with mass $m_{\phi}$.}
\label{tabs7}
\end{table}
Finally, the
non-analytic functions ${\cal L}$'s, ${\cal J}$'s, ${\cal K}$'s , 
${\cal R}$'s, ${\cal T}$'s and ${\cal S}$'s in above equations 
are defined in section \ref{IST}. 
Employing the same argument as one did in deriving the isospin changing
pseudoscalar form factors, one arrives at 
a similar expression for the strangeness changing 
pseudoscalar form factor 
\begin{eqnarray}
G_{P,B'B}(q^2)\,& = &\,(m_B+m_{B'})^2
 \Big(\,\frac{f_{K}/f}{q^2-m^2_{K}}\,G_{A,B'B}(0)\sqrt{Z_{K}}\, 
\,- \, \frac{1}{3}\langle r^{2}_{B'B} \rangle\,\Big)\,,
\label{pseudostrange}
\end{eqnarray}  
where in this NLO expression, the axial charge $G_{A,B'B}(0)$,
kaon mass $m_{K}$ and kaon decay constant $f_{K}$ are taken to be 
their physical values and $Z_{K}$ is the kaon wavefunction renormalization
which is shown in the Appendix. In fitting lattice
data to Eq.~ (\ref{pseudostrange}), one would thus use the lattice measured 
values for $G_{A,B'B}(0)$, $m_{K}$ and $f_{K}$.  The final 
term in the pseudoscalar form factor is $\langle r^{2}_{B'B} \rangle$, 
which is the strangeness changing axial radius. 
As has been shown in section \ref{IST}, 
the simple structure of the psuedoscalar form factor at NLO
in both \CPT\ and \PQCPT\, allows one perform an approximate check
of the Goldberger-Treiman relation. Here the residue of the pseudoscalar
form factor at the kaon pole is proportional to the kaon-baryon-baryon
coupling $G_{K B'B}$ thus the pseudoscalar form factor provides
an indirect and simple method to investigate the Goldberger-Treiman
relation on the lattice.

\section{Discussion}
\label{discussion} 
Above we have calculated the full set of flavor-changing
axial-current matrix elements of the hyperons.
The expressions will be useful for the study of the 
chiral and momentum behavior of hyperon axial form factors 
using lattice QCD. 

Considering the axial charges of hyperons and of their 
transitions, there are only two parameters which survive 
the chiral limit, namely $D$ and $F$. 
As there are eight such charges, there are six relations 
between them. Focusing just on the isospin and strangeness
transitions individually, we have four of the six relations
\begin{eqnarray}
&& 
g_{NN} + g_{\Xi \Xi} - g_{\Lambda \Sigma} = 0\,, 
\label{eq:firstsetone}
\\
&& 
g_{NN} - g_{\Xi \Xi} - g_{\Sigma \Sigma} = 0\,,
\\
&&
g_{N \Lambda} + g_{\Lambda \Xi} - 3 ( g_{\Sigma \Xi} - g_{N \Sigma} )
= 0\,,
\\
&&
g_{N \Lambda} - g_{\Lambda \Xi} - g_{N \Sigma} - g_{\Sigma \Xi}
= 0\,
\label{eq:lastsetone}
.\end{eqnarray}
Combining isospin and strangeness transitions
together, we arrive at the final two relations
\begin{eqnarray}
&& 
g_{N \Lambda} + g_{\Lambda \Xi} - 3 g_{\Sigma \Sigma} = 0\,,
\label{eq:firstsettwo}
\\
&&
g_{N \Lambda} - g_{\Lambda \Xi} - g_{\Lambda \Sigma} = 0\,
\label{eq:lastsettwo}
.\end{eqnarray}
These relations hold in $SU(3)$ $\chi$PT, as well as $SU(6|3)$ 
PQ$\chi$PT. Of course chiral corrections modify these relations
and the ``$0$" should be interpreted as 
$\mathcal{O}( m_\phi^2 / \Lambda_\chi^2)$. 
The expressions derived above in section \ref{IST} and \ref{SCT} 
provide these $\mathcal{O} (m_\phi^2 /\Lambda_\chi^2)$
corrections which are generally linear combinations of non-analytic loop
contributions and unknown local counterterms.

The relations in Eqs.~\eqref{eq:firstsetone}--\eqref{eq:lastsetone}
actually apply not only to the axial charges, but 
to the respective hyperon transition matrix elements as a whole. 
The pseudoscalar form factors do 
not satisfy Eqs.~\eqref{eq:firstsettwo} and \eqref{eq:lastsettwo}
due to the difference in pion versus kaon poles. The axial 
form factors, by contrast, satisfy these latter two relations. 

Apart from the axial couplings $D$, $F$, $\mathcal{C}$, $\mathcal{H}$,
and meson masses, the axial charges depend on six (eight) unknown 
parameters in $\chi$PT (PQ$\chi$PT). This lack of predictive power
historically has been overlooked by supposing, what has been termed,
the formal dominance of chiral logarithms. One can do slightly 
better and attempt model estimation of these parameters, but 
the fact stands that most chiral analyses have seriously lacked 
the ability to make complete predictions. This is the great
advantage of lattice QCD calculations, which have promise to determine 
the complete information about the low-energy constants.

Despite the absence of knowledge concerning the eight coupling constants 
in the NLO current Eq.~(\ref{eq:axcts}), we can make two non-trivial predictions
at next-to-leading order by eliminating the local terms. 
The leading local terms are cancelled in the relations Eqs.~\eqref{eq:firstsetone}--\eqref{eq:lastsettwo}.
Combinations of these relations can be used to eliminate the next-to-leading order 
local terms. While there are seven contributing terms, only five of them, 
$b_1, \ldots, b_5$ 
contribute to the relations. 
Thus there must exist at least one combination of Eqs.~\eqref{eq:firstsetone}--\eqref{eq:lastsettwo}
that is independent of the 
$b_i$. 
Rather fortuitously there are two such 
non-trivial combinations
of axial charges which are independent of these 
couplings, \emph{viz}.
\begin{eqnarray}
\Delta g 
&\equiv& 
2 g_{NN} - g_{N \Lambda} - g_{N \Sigma} - g_{\Lambda \Sigma}
- g_{\Sigma \Sigma} + 2 g_{\Sigma \Xi}\,,
\\
\Delta G
&\equiv&
2 g_{NN} + 2 g_{\Xi\Xi} - 2 g_{\L \S}
+ 
g_{N\S} + g_{\L \Xi} + g_{\S \Xi} - g_{N \L}
.\end{eqnarray}
These relations are independent of NLO counterterms in both 
PQ$\chi$PT
and 
$\chi$PT. 
The expressions for $\D g$ and $\D G$ in 
$\chi$PT 
are%
\footnote{
Partially quenched expressions can be obtained 
from Eqs.~\eqref{eq:deltag} and \eqref{eq:deltaG}
under the replacements:
$\mathcal{G} [\cL] \to \mathcal{G}^{PQ} [\cL, \cR]$,
$\mathcal{G} [\cK] \to \mathcal{G}^{PQ} [\cL, \mathcal{S}]$,
and
$\mathcal{G} [\cJ] \to \mathcal{G}^{PQ} [\cL, \cT]$.
The partially quenched functional is given by
\begin{equation}
\mathcal{G}^{PQ} [A,B]
=
2 A_\pi - 4 A_K + 2 A_{\eta_s} + 2 B_{\eta_u \eta_u} - 4 B_{\eta_u \eta_s} + 2 B_{\eta_s \eta_s}
\notag
,\end{equation}
for any function 
$A_\phi = A(m_\phi,\D,\mu)$
and hairpin function
$B_{\eta_{\phi}\eta_{\phi'}} = B(\eta_{\phi}\eta_{\phi'},\D,\mu)$.
There is no scale dependence in 
$\mathcal{G}^{PQ} [A,B]$.
}
\begin{figure}
\begin{center}
\includegraphics[width=0.4825\textwidth]{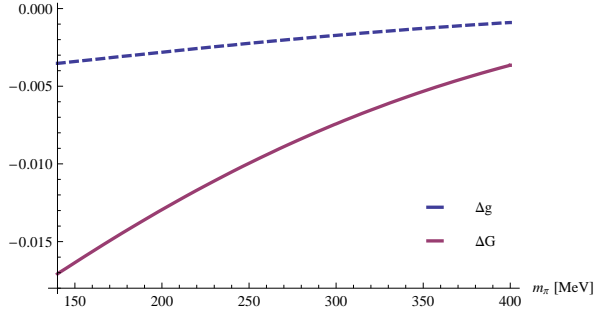}
\end{center}
\caption{Plot of $\Delta g$ and $\D G$ as a function of the pion mass $m_{\pi}$.} 
\label{fig3}
\end{figure}
\begin{eqnarray}
(4 \pi f)^2 \, \Delta g &=& \frac{2}{3} \Big( D^3 + 5 D^2 F
- 6 D F^2 - 6 F^3 \Big) \, 
\mathcal{G} [\cL]
\nonumber \\
&& -  \mathcal{C}^2 \Big[ \frac{10}{81} \mathcal{H} + \frac{1}{6}(D + F) \Big]\, 
\mathcal{G} [\cJ]
-  \frac{2}{9} \mathcal{C}^2 (D -F)\, 
\mathcal{G} [\cK]\,,
\label{eq:deltag}
\\
(4 \pi f)^2 \, \Delta G &=& \frac{4}{3} D ( D^2  - 6 F^2)\,
\mathcal{G} [\cL]
- \frac{1}{3} D \mathcal{C}^2 \,
\mathcal{G} [\cJ]
+ \frac{4}{3} F \mathcal{C}^2 \,  
\mathcal{G} [\cK]
\label{eq:deltaG}
\,,\end{eqnarray}
where the Gell-Mann Okubo linear combination functional is defined by
\begin{equation}
\mathcal{G}[A] = A_{\pi} - 4 A_{K} + 3 A_{\eta}
\,,\end{equation}
for any function $A_\phi = A(m_\phi,\D,\mu)$. 
The quantities $\D g$ and $\D G$ allow one to test chiral logarithms directly. 
Accordingly these relations only superficially have scale 
dependence, and upon the limit of $SU(3)_V$ symmetry, they vanish. 
With these symmetry breaking relations, we have separated out the 
short distance physics and hence isolated long-distance 
chiral corrections. To obtain values for $\D g$ and $\D G$, as well as the curve shown in 
Fig.~\ref{fig3}, we have fixed the strange quark mass at its physical 
value, and used the $SU(6)$ values for the axial couplings: $D = 3/4$,
$F = 2/3 D$, $\mathcal{C} = - 2 D$, and $\mathcal{H} = - 3 D$. 
In particular, we find
$\Delta g = -0.0035$ 
and
$\Delta G  = -0.017$ 
at physical pion mass $m_{\pi}\,=\,140\, {\text{MeV}}$.

Four further non-trivial relations, analogous to 
$\Delta g$ and $\D G$ 
above,
exist when one includes charges
arising from the strangeness axial current
$\ol s \gamma_\mu \gamma_5 s$. 
These matrix elements have been determined at zero momentum 
transfer in 
PQ$\chi$PT~%
\cite{will:2004}.
Determination of these charges requires the calculation of disconnected
operator contractions on the lattice. As such contributions are notoriously
difficult to determine, 
we leave it to future work to deduce the remaining symmetry
breaking relations.

\begin{acknowledgments}
We thank G. Colangelo for discussions. This work is supported in part by 
the U.S.\ Dept.~of Energy, Grant Nos.\ DE-FG02-05ER41368-0 and 
DE-FG02-93ER40762 (B.C.T.) and by the Schweizerischer Nationalfonds (F.-J.J.). 
\end{acknowledgments}

\appendix

\section*{Wave Function Renormalization} \label{s:wfn}

In this appendix, we list the necessary wave function renormalization and
meson $Z-$factors appearing in the calculations.
For the baryons, we have \cite{MSavage:2002.1}
\begin{eqnarray}
Z_{B} \,=\,1-\frac{1}{16\pi^2f^2}\Bigg( \sum_{\phi}{\cal A}_{\phi}{\cal L}_{\phi} 
+\sum_{\phi,\phi'}{\cal A}_{\phi\phi'}{\cal R}_{\phi\phi'}
+{\cal C}^2\Big(\sum_{\phi}{\cal B}_{\phi}{\cal J}_{\phi} 
+\sum_{\phi,\phi'}{\cal B}_{\phi\phi'}{\cal
  T}_{\phi\phi'}\Big)\Bigg)
\end{eqnarray}
where the coefficeint ${\cal A}_{\phi}$, $\bar{{\cal A}}_{\phi,\phi'}$, 
${\cal B}_{\phi}$, $\bar{{\cal B}}_{\phi\phi'}$ and are given in 
Tables \ref{tabwr1}, \ref{tabwr2}. For the mesons, one has
\begin{equation}
Z_{\varphi} \,=\, 1 \,+\,\frac{2}{3}\Big(\frac{1}{16\pi^2f^2}
\sum_{\phi}C_{\phi}{\cal L}
(m_{\phi},\mu)\,\Big)-\frac{2}{\lambda}\Big( (2m^2_{jj}+m^2_{rr})\alpha_4
+ w_{\varphi}\alpha_5\Big),
\end{equation}
where $\alpha_4$ and $\alpha_5$ are the low energy constants that
appear in ${\cal L}^{4}$ in the meson sector. Further,
for $\varphi \,=\,\pi$, $w_{\pi} \,=\,m^2_{uu} $ and one 
uses Table \ref{tabi1} for $C_\phi$ while for $\varphi \,=\, K$, 
$w_{K} \,=\,m^2_{us}$ and one uses Table \ref{tabs1} for 
$C_\phi$. Finally the non-analytic functions
${\cal J}$'s and ${\cal L}$'s are given in section \ref{IST}.

\begin{table}[!ht]
\begin{center}
\begin{tabular}{cccc|c}
\multicolumn{4}{c|}{${\cal A}_{\phi}$}&
{$\bar{{\cal A}}_{\phi\phi'}$}\tabularnewline 
\hline 
&
$\eta_{u}$ & $\eta_{s}$ & $ us$ & $\eta_{u}\eta_{u}$\tabularnewline
$N$ & $-4D(D-3F)$ & $0$ & $0$ &$3(D-3F)^2$\tabularnewline
$\Lambda$ & $-\frac{2}{3}D^2+8DF-6F^3$ & $0$ & 
$-\frac{10}{3}D^2+4DF+6F^2 $ & $\frac{4}{3}(2D-3F)^2$\tabularnewline
$\Sigma$ & $2(3F^2-D^2)$ & $0$ & $-2(D^2-6DF+3F^2)$ & $12F^2$\tabularnewline
$\Xi$ & $0$ & $2(3F^2-D^2)$ & $-2(D^2-6DF+3F^2)$ & $3(D-F)^2$\tabularnewline
\hline 
\hline
&
$uj$ & $ur$ & $js$ & $\eta_{u}\eta_{s}$\tabularnewline
$N$ & $10D^2-12DF+18F^2$ & $5D^2-6DF+9F^2$ & $0$ & $0$\tabularnewline
$\Lambda$ & $\frac{28}{3}D^2-16DF+12F^2$ & 
$\frac{14}{3}D^2-8DF+6F^2$ & $\frac{2}{3}(D+3F)^2$ &$-\frac{4}{3}(2D^2+3DF-9F^2)$\tabularnewline
$\Sigma$ & $4D^2+12F^2$ & 
$2D^2+6F^2$ & $6(D-F)^2$ & $12F(F-D)$\tabularnewline
$\Xi$ & $6(D-F)^2$ & $3(D-F)^2$ & $4(D^2+3F^2)$ & $12F(F-D)$\tabularnewline
\hline 
\hline
&
$sr$ & $$ & $$ & $\eta_{s}\eta_{s}$\tabularnewline
$N$ & $0$ & $$ & 
$$ & $0$\tabularnewline
$\Lambda$ & $\frac{1}{3}(D+3F)^2$ & $$ & 
$$ & $\frac{1}{3}(D+3F)^2$\tabularnewline
$\Sigma$ & $3(D-F)^2$ & 
$$ & $$ & $3(D-F)^2$\tabularnewline
$\Xi$ & $2(D^2+3F^2)$ & $$ &
$$ & $12F^2$\tabularnewline
\end{tabular}
\end{center}
\caption{The coefficients ${\cal A}_{\phi}$ and $\bar{{\cal A}}_{\phi\phi'}$ 
in PQ$\chi$PT for the wave function renormalization. 
The ${\cal A}_{\phi}$ and $\bar{{\cal A}}_{\phi\phi'}$ coefficients are
categorized as in Table \ref{tabi2}.}
\label{tabwr1}
\end{table}

\begin{table}[!ht]
\begin{center}
\begin{tabular}{cccccccc|ccc}
\multicolumn{8}{c|}{${\cal B}_{\phi}$}&
\multicolumn{3}{c}{$\bar{{\cal B}}_{\phi\phi'}$}\tabularnewline
\cline{1-8} 
\hline 
&
$\eta_{u}$&
$\eta_{s}$&
$us$&
$uj$&
$ur$&
$sj$&
$sr$&
$\eta_{u}\eta_{u}$&
$\eta_{u}\eta_{s}$&
$\eta_{s}\eta_{s}$\tabularnewline
$N$&
$2$&
$0$&
$0$&
$2$&
$1$&
$0$&
$0$&
$0$&
$0$&
$0$\tabularnewline
$\Lambda$&
$1$&
$0$&
$1$&
$2$&
$1$&
$0$&
$0$&
$0$&
$0$&
$0$\tabularnewline
$\Sigma$&
$\frac{1}{3}$&
$0$&
$\frac{5}{3}$&
$\frac{2}{3}$&
$\frac{1}{3}$&
$\frac{4}{3}$&
$\frac{2}{3}$&
$\frac{2}{3}$&
$-\frac{4}{3}$&
$\frac{2}{3}$\tabularnewline
$\Xi$&
$0$&
$\frac{1}{3}$&
$\frac{5}{3}$&
$\frac{4}{3}$&
$\frac{2}{3}$&
$\frac{2}{3}$&
$\frac{1}{3}$&
$\frac{2}{3}$&
$-\frac{4}{3}$&
$\frac{2}{3}$\tabularnewline
\end{tabular}
\end{center}
\caption{The coefficients ${\cal B}_{\phi}$ and $\bar{{\cal B}}_{\phi\phi'}$ 
in PQ$\chi$PT for the wave function renormalization. 
The ${\cal B}_{\phi}$ and $\bar{{\cal B}}_{\phi\phi'}$ coefficients are
categorized as in Table \ref{tabi2}.}
\label{tabwr2}
\end{table}

\end{document}